\documentclass[a4paper,11pt]{article}
\usepackage{jheppub}
\usepackage{amsmath}
\usepackage{amsthm}
\usepackage{amsfonts}
\usepackage{amssymb}
\usepackage{textcomp}
\usepackage{graphicx}
\usepackage{bm}
\usepackage{color}
\usepackage{verbatim}
\usepackage{subcaption}
\usepackage{graphicx,color}
\usepackage{epsfig}

\definecolor{orange}{rgb}{1,0.5,0}
\definecolor{col1}{RGB}{153, 52, 121}
\definecolor{dgreen}{rgb}{0,0.55,0}
\definecolor{pink}{rgb}{1,0.08,0.58}

\usepackage{amsfonts,amsmath,amssymb}

\newcommand{\la}{\langle}
\newcommand{\ra}{\rangle}

\newcommand{\rar}{\rightarrow}

\theoremstyle{definition}

\begin{document}

 \title{
 \Huge Coherent vs incoherent transport \\
 in holographic strange insulators \color{black}}
 
 \author[a]{Tomas Andrade}
 
 \author[b,c]{Alexander Krikun\footnote{https://orcid.org/0000-0001-8789-8703}}
 
\affiliation[a]{
Departament de F{\'\i}sica Qu\`antica i Astrof\'{\i}sica, Institut de
Ci\`encies del Cosmos, Universitat de
Barcelona, \\ Mart\'{\i} i Franqu\`es 1, E-08028 Barcelona, Spain}

\affiliation[b]{Instituut-Lorentz for Theoretical Physics, $\Delta ITP$, Leiden University, Niels Bohrweg 2, Leiden 2333CA, The Netherlands} 
\affiliation[c]{Institute for Theoretical and Experimental Physics (ITEP)\footnote{On leave from}, \\ B. 
 Cheryomushkinskaya 25, 117218 Moscow, Russia }

\emailAdd{tandrade@icc.ub.edu }
\emailAdd{krikun@lorentz.leidenuniv.nl}

\abstract{
Holographic strange metals are known to have a power law resistivity rising with temperature, which is 
reminiscent of the strange metal phases in condensed matter systems. 
In some holographic models, however, the exponent of the power law in the resistivity can be negative. 
In this case one encounters phases with diverging resistivity at zero temperature: holographic strange insulators. 

These states arise as a result of translational symmetry breaking in the system, which can either be strong explicit and relevant in the IR, or spontaneous, but pinned by a small explicit source. In some regards, one can associate these two classes to the normal band insulators 
due to the strong ionic potential, and Mott insulator due to the commensurate lock in of the charge density wave. 

We study different features of these classes on the explicit example of a holographic helical model with homogeneous Bianchy 
VII type translational symmetry breaking, and uncover the main mechanisms underlying transport in these two cases. 
We find that while transport in the explicit relevant case is governed by the incoherent conductivity, in the pinned spontaneous case the leading contribution comes from the coherent part.
}

\maketitle

\section{Introduction}

The interplay between explicit and spontaneous translational symmetry breaking has recently become a major subject in the field of Holographic Duality and its applications to Condensed Matter. It is indeed relevant because any condensed matter system is realized on top of an ionic crystal lattice, and its (explicit) effect has to be well understood. 
Moreover, some of the most interesting condensed matter systems, including high temperature superconductors, exhibit 
phases with spontaneously generated spatial orders, as charge, spin and pair density waves. 
It has recently been discussed \cite{Mott} that the Mott insulating state in the parent compound of the cuprate high temperature superconductors can be understood as the spontaneous charge density wave (CDW) pinned by the commensurate crystal lattice. 
Furthermore, the fluctuating CDW has been proposed as the physical mechanism of bad metallic behavior \cite{Delacretaz:2016ivq,Delacretaz:2017zxd}. It is therefore important to develop our understanding of the features of translation symmetry breaking in holographic models. 

Although the physics of translation symmetry breaking (TSB) in holographic models is in many ways similar to the one in conventional systems, some of its features are unusual. The well known example is the absence of a gapped insulating state in holographic models \cite{Grozdanov:2015qia}. In this work we will focus on the details of the ``gapless insulating'', or rather semimetallic, states, which one can obtain in holography. 
Similarly to the ``holographic strange metal'' phase, which is argued to have many features in common with the real strange metal states observed in strongly correlated condensed matter systems \cite{Davison:2013txa, Zaanen:2015oix}, the holographic ``strange insulating'' states exhibit resistivity which behaves as a power law of the temperature. 
The only difference is that while in a metal state the resistivity vanishes at zero temperature, in the insulating state it diverges having a negative power law exponent. Interestingly, one can achieve this type of behavior in two seemingly different ways:
one can either introduce strong explicit TSB, which becomes relevant at the horizon and therefore changes the zero temperature ground 
state of the model \cite{Donos:2012js,Donos:2014uba,Rangamani:2015hka}, or one can consider a model with spontaneous TSB and 
introduce a weak explicit source to pin down the spontaneous structure \cite{Andrade:2017cnc, Jokela:2017ltu, Mott}. 
In both cases one arrives at a holographic ``strange insulating'' state at small enough temperature. Interestingly, the analogues of these two classes can be found in conventional Condensed Matter Theory. 
We can think about the usual band insulator as being induced solely by the strong ionic lattice potential, i.e., even the weakly interacting electrons would be gapped due to the shape of the energy bands, induced by the periodic potential. 
On the other hand, the Mott insulator can be seen as a pinned crystal of electrons: the Coulomb repulsion now plays the crucial role, which would provide the stiffness to the spontaneous Wigner crystal even in the case when the ionic lattice is absent. These analogies motivate us to call these states insulators due to the fundamental TSB mechanisms giving rise to these states, even though from the phenomenological perspective these should rather be called semimetals due to their nonvanishing conductivity at any finite temperature.
Given these similarities, it is interesting to perform a more detailed comparison between the two insulating configurations 
arising in the holographic approach.  

From the bulk perspective, the common feature to both classes is that in either case TSB is relevant in the IR and the near horizon geometry is substantially modified by the spatially modulated structures. It is therefore intuitively clear that the DC conductivity will be sensitive to this modification \cite{Donos:2014cya}. One would also expect the incoherent conductivity to play an important role in the 
defining the features of the low temperature state \cite{Davison:2015bea}, since by its definition it is not associated with the momentum drag \cite{Davison:2015taa} and therefore should not be suppressed significantly by the TSB.
This makes our study more interesting and relevant to experiments on strongly correlated electron systems \cite{ando1995logarithmic}, which exhibit unusual resistivity behavior in low temperature pinned CDW states. 
The reason for this is that the incoherent conductivity is a genuinely holographic feature, which is absent in the conventional Condensed Matter models. 
It is commonly understood as arising from the quantum critical subpart of the strongly interacting quantum system having a holographic dual \cite{Davison:2015bea, Krikun:2018agd}.
Its direct observation in experiments would serve as a smoking gun for the existence of such a quantum critical sector in  real strongly correlated materials.
On the other hand, it has been shown in \cite{Delacretaz:2016ivq} that an extra contribution to the conductivity in the pinned CDW 
state may come from the ``phase relaxation rate'', which can be understood as a feature of the coherent subpart of the system, related to the approximate conservation of momentum and fluctuations of the weak spontaneous order parameter. 
With this in mind, it is very interesting to disentangle the incoherent and, more conventional, coherent contributions in 
the holographic observables and explore their roles in the two classes of insulating states outlined above. 

In what follows we will put this intuition on a solid quantitative ground by considering the two ``strange insulating'' states in 
a concrete holographic model and studying their similarities and differences from complimentary points of view. 
We will use a particular holographic model with Bianchy VII helical symmetry \cite{Nakamura:2009tf, Ooguri:2010kt,Donos:2012js,Donos:2012wi,Donos:2014oha}. As the linear axion \cite{Andrade:2013gsa, Baggioli:2016pia}, Q-lattice\cite{Donos:2013eha, Donos:2014uba} and their generalizations, the model belongs to the class of the so-called ``holographic homogeneous lattices'': it realizes the translational symmetry breaking, but the presence of an extra internal group which can be used to restore the symmetry allows one to treat the equations of motion in a homogeneous manner.\footnote{A similar family of models is obtained by giving an explicit mass term to the graviton \cite{Alberte:2017oqx,Alberte:2017cch}}This feature leads to a tremendous technical simplification, since one has to deal only with ordinary differential equations instead of partial ones and, as we will discuss below, some powerful analytical techniques to treat the near horizon features become available. However, the homogeneous lattices are known to introduce some conceptual subtleties in exchange for this technical simplicity \cite{Andrade:2015iyf, Musso:2018wbv} therefore we have to be careful with this choice.

The helical model which we will use is particularly suitable for our studies since it is the only one from the family of ``holographic homogeneous lattices'', which provides a clean and well studied mechanism for spontaneous translational symmetry breaking -- a crucial part of the phenomenon under investigation. One can study the critical temperature for the phase transition \cite{Nakamura:2009tf}, find the unstable modes in the linearized spectrum of the normal homogeneous phase \cite{Ooguri:2010kt}, follow the thermodynamic potential of the new symmetry broken ground state \cite{Donos:2012wi}, identify the order parameter, thermodynamically preferred value of the helical pitch in the broken phase and the stiffness (Young modulus) of the spontaneously generated spatial structure \cite{Andrade:2017cnc}. All these features studied previously in the literature follow the common physical intuition behind the spontaneous crystallization, therefore it makes us confident that symmetry breaking pattern we are going to study is not pathological in any way. 

An explicit symmetry breaking can be introduced in exactly the same framework \cite{Donos:2012js}. In our setup, the explicit 
source couples to a separate field in the bulk, what makes the distinction between explicit and spontaneous mechanisms of TSB unambiguous \cite{Andrade:2017cnc}. Importantly for the present study, unlike linear axions, in the helical model the explicit breaking can be relevant at the horizon, giving rise to a distinct ``anisotropic'' horizon geometry \cite{Donos:2014oha}, which is algebraically insulating and will be the focus of the present study. 

We outline the details of the model and discuss the features of the ``explicit'' and ``pinned spontaneous'' insulating states together with their zero temperature fixed points in Sec.\,\ref{sec:model}. 
In Sec.\,\ref{sec:Inc_conductivity} we introduce the conductivity matrix and, in particular, identify the expression for the incoherent conductivity at finite frequency. 
We first study the DC response in Sec.\ref{sec:DC_conductivity}. We extract conductivities from the near horizon data, check that the two classes of solutions are indeed gapless insulating, and match the analytically predicted low temperature scalings. At finite temperature, 
we clearly observe that the incoherent transport dominates in the relevant explicit case while it is subleading in the pinned spontaneous one.
Then in Sec. \,\ref{sec:AC_conductivity} we turn to the AC response. We evaluate it numerically by solving the linearized equations of motion, and show that the incoherent conductivity does indeed remove the peak from the AC conductivity. 
We follow the changes in the conductivity as the temperature is lowered and match the results with the calculation of 
the quasinormal modes. This more detailed analysis shows that the coherent subsystem remains in the pinned spontaneous insulator even at low temperatures.
In Sec.\,\ref{sec:Omega} we analyze the coherent contributions in the two cases by fitting the lineshapes to the hydrodynamic predictions and we conclude in Sec.\,\ref{sec:Conclusion}. The Appendices are devoted to the details of the numerical scheme we use to obtain background solutions (App.\,\ref{app:background}), zero temperature asymtpotic analysis (App.\,\ref{app:zero_T}), derivation of the DC formulas from near horizon data (App.\,\ref{app:DC_formulas}) and the solution to the linearized equations of motion for AC conductivities and quasinormal modes (App.\,\ref{app:AC_conductivities}).

\section{\label{sec:model}Insulating groundstates of the helical model}

\begin{figure}[t]
\centering
\includegraphics[width=0.49 \linewidth]{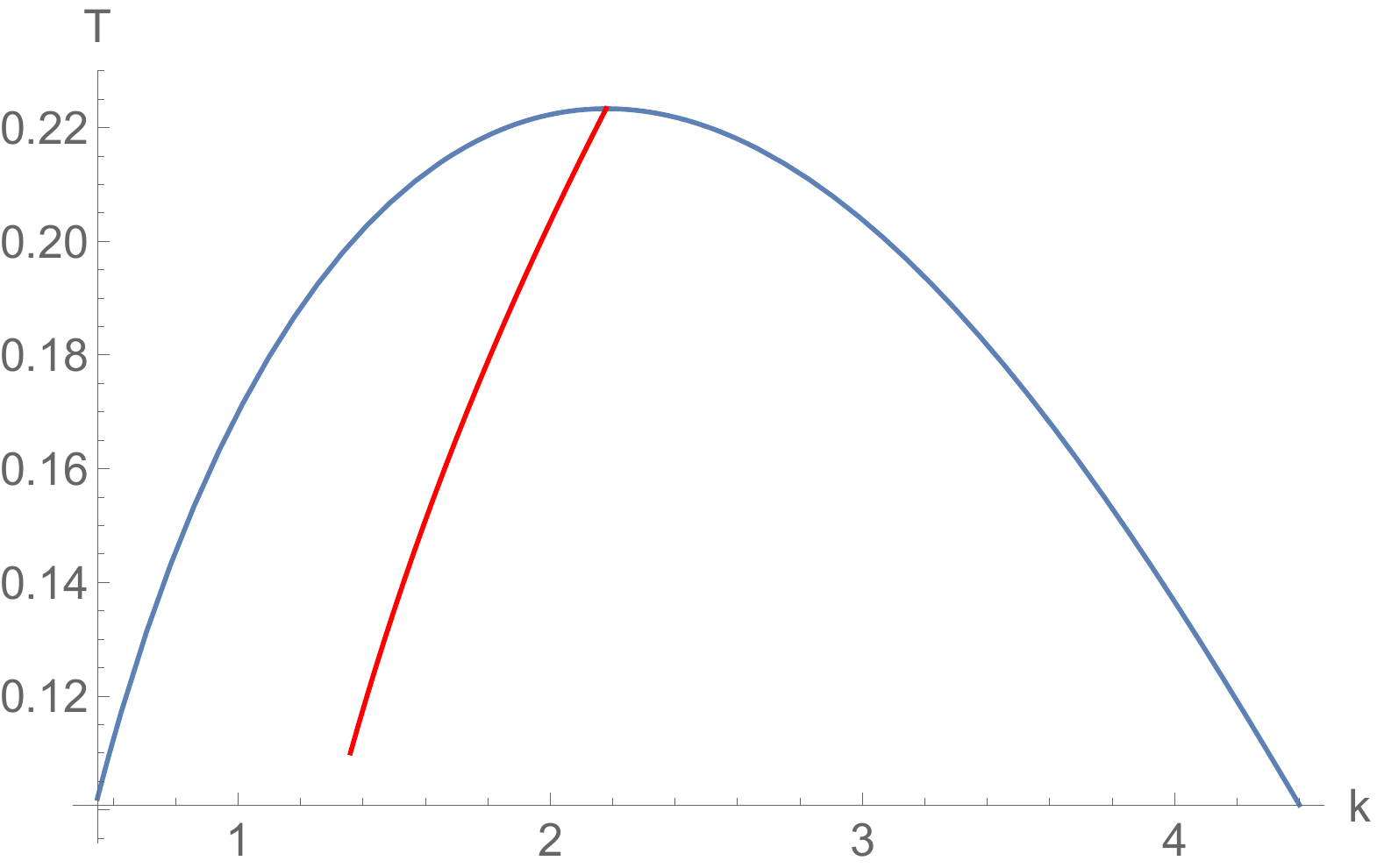}
\includegraphics[width=0.49 \linewidth]{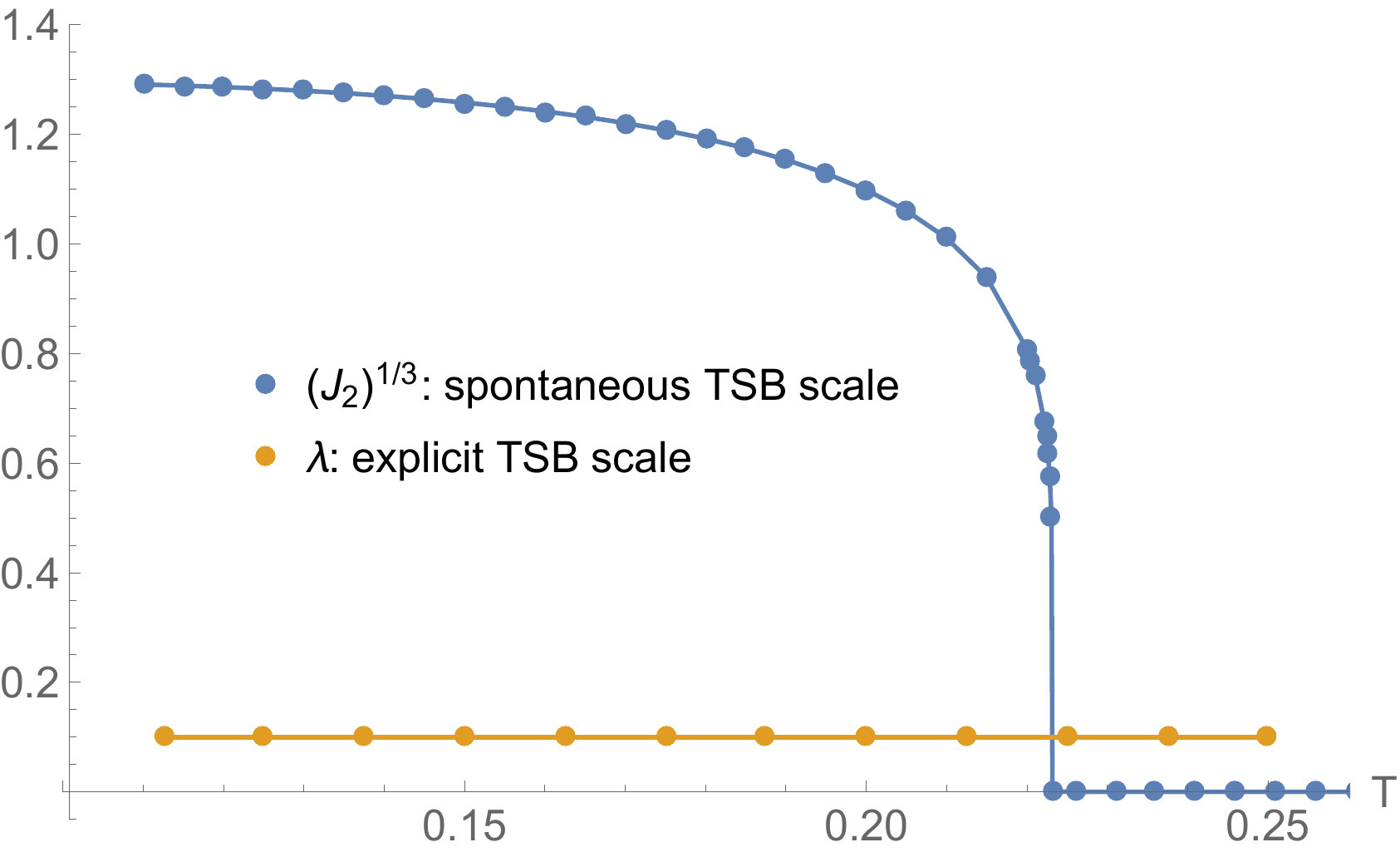}
\caption{\textbf{Spontaneous translational symmetry breaking.} \textit{Left:} Bell curve of marginal modes for $\gamma=3, \kappa = 0$ (blue), showing the temperatures at which the instability with given momentum $k$ develops in the Reissner-Nordstrom background. This instability drives the spontaneous phase transition. The endpoint of the transition, the thermodynamically preferred non-linear solution has the momentum show by the red line.
\textit{Right:} temperature dependence of the order parameter -- the expectation value for the spontaneous helical current $J_2$ as compared with the scale of the explicit pinning source $\lambda = 0.1 \mu$. In all cases studied henceforth the pinning explicit scale is small compared to the spontaneous one.}%
\label{fig:bell}
\end{figure}

We study the holographic model in 5-dimensional ($x_\mu = \{t,x,y,z,r\}$) bulk with dynamical gravity, an Abelian gauge 
field $A_\mu$, dual to the chemical potential and an auxiliary vector field $B_\mu$, which will be used to source the explicit 
translational symmetry breaking. The action reads
\begin{equation}
\label{eq:action}
 	S = \int d^5 x \sqrt{- g} \left( R  - 2 \Lambda - \frac{1}{4} F^2 - \frac{1}{4} W^2  \right) - 
 	 \frac{\gamma}{6} \int  A \wedge F \wedge F - \frac{\kappa}{2} \int B \wedge F \wedge W,
\end{equation} 
where $\Lambda=-6$ and $F \equiv dA$, $W\equiv dB$ -- the field strength tensors. We set the mass of the $B$ field to zero, but this
will not play an important role in our results. 
This model has been studied in depth in the holographic context. In \cite{Nakamura:2009tf} it has been shown that the model with $\kappa = 0$ in absence of the $B$ field develops a dynamical instability (see Fig.\,\ref{fig:bell}) due to the Chern-Simons term $\gamma$, which breaks the $x$-translation and $(y,z)$-rotation symmetries down to a diagonal subgroup. The endpoint of the instability was studied in \cite{Ooguri:2010kt, Donos:2012wi}, where it was shown that the new groundstate does indeed break the translations spontaneously and has a Bianchy~VII type geometry described in terms of the helical forms
\begin{align}
\label{equ:helical_forms}
\omega^{(k)}_1 & = dx \\
\omega^{(k)}_2 & = \cos (k x) dy - \sin(k x) dz \\
\omega^{(k)}_3 & = \sin (k x) dy + \cos(k x) dz. 
\end{align}

A different type of solutions has been explored in \cite{Donos:2012js, Donos:2014oha} in the model with $\gamma=0$, but with finite boundary value of the helical mode of the $B$-field: $B\big|_{u\rar0} = \lambda \omega_2$. In these models, translation symmetry is broken explicitly by the external source $\lambda$. The momentum is no longer conserved and therefore at finite charge density these models display finite resistivity due to momentum dissipation. At finite $\kappa$ there are two distinct groundstates, depending on the value of the explicit source and helical pitch $k$: the metallic phase, with vanishing resistivity at zero temperature, and the ``insulating'' phase, defined as the one with diverging resistivity. 

More recently in \cite{Andrade:2017cnc} we studied the model with both spontaneous and explicit symmetry breaking mechanisms and found that when the Goldstone mode arising due to spontaneous breaking gets pinned by explicit source, the resulting state becomes 
again ``insulating'' in a similar way: the resistivity grows towards small temperature.  

Both types of solutions are captured by the ansatz
\begin{gather}
\label{ansatz}
ds^2 = \frac{1}{u^2}\left[ - T(\!u) f(\!u) dt^2 + \frac{U(\!u) du^2}{f(\!u)} + W_1(\!u) \omega_1^2 + W_2(\!u) \big(\omega_2+
Q(\!u) dt\big)^2 + W_3(\!u) \omega_3^2 \right] \\
A = A_t(\!u) dt + A_2(\!u) \omega_2, \qquad  B = B_t(\!u) dt + B_2(\!u) \omega_2,
\end{gather}
where in the pure spontaneous case $B_t, B_2=0$ and in the pure explicit case $A_2,Q = 0$.
We keep the system at finite chemical potential by fixing the boundary condition $A_t = \mu + O(u^2)$ near the 
conformal boundary $u=0$. 
We choose 
$f = (1-u^2)(1 + u^2 - u^4 \mu /3 )$, so that the Reissner-Nordstr\"om (RN) solution has the simple form $T = U= W_i = 1$, 
$Q=0$, $A_t = \mu(1 - u^2)$, $A_2 = B_\mu = 0$. In the numerical calculations we fix the gauge redundancy of the ansatz \eqref{ansatz} by 
using the DeTurk trick \cite{Headrick:2009pv,Adam:2011dn,Wiseman:2011by} (see also Appendix\,\ref{app:background}). 
This implies that the temperature of the solutions is given by 
\begin{equation}
	\frac{T}{\mu} = \frac{6 - \mu^2}{6 \pi \mu}.
\end{equation}
In what follows we will measure all the dimensionful quantities in units of $\mu$.

\begin{figure}[t]
\centering
\begin{minipage}{0.44 \linewidth}
\includegraphics[width=1 \linewidth]{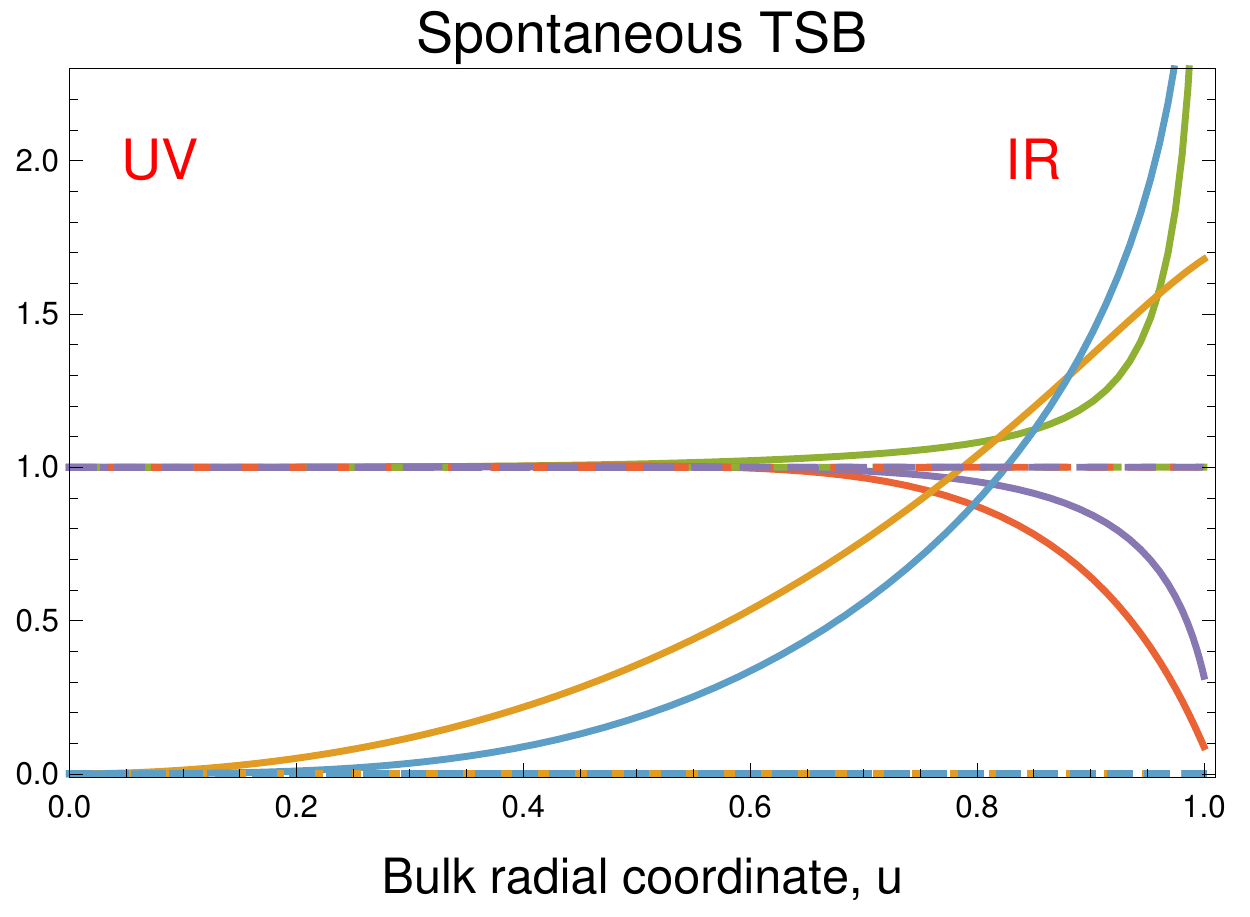}
\end{minipage}
\begin{minipage}{0.1 \linewidth}
\includegraphics[width=1 \linewidth]{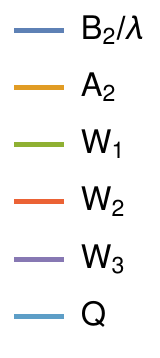}
\end{minipage}
\begin{minipage}{0.44 \linewidth}
\includegraphics[width=1 \linewidth]{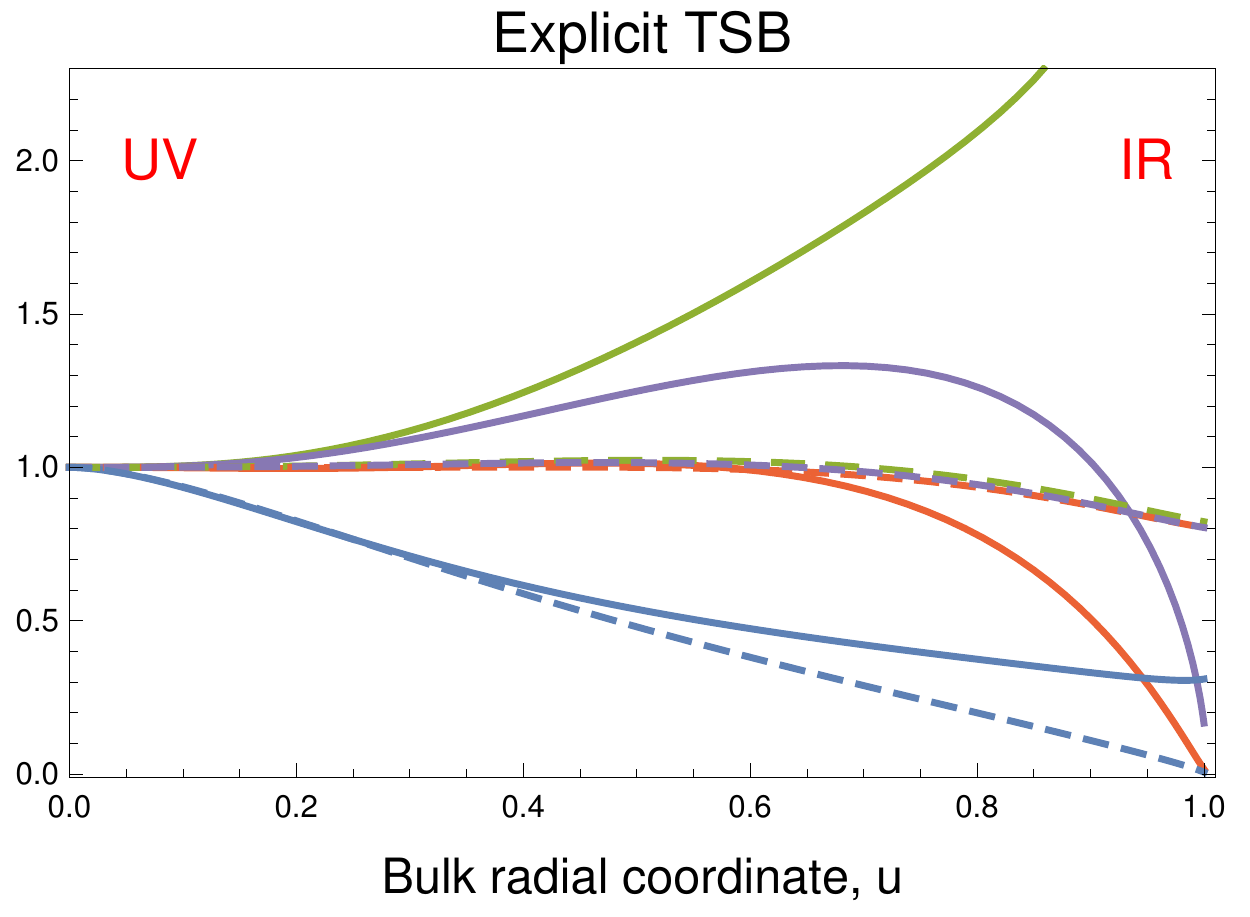}
\end{minipage}
\caption{\textbf{Background profiles} in cases of spontaneous (left) and explicit (right) translation symmetry breaking. Solid lines show the cases when the TSB is relevant at the horizon, and we use dashed for the irrelevant ones. It is clearly seen that in the relevant case the field $W_1$ diverges at the horizon, while $W_2$ and $W_3$ vanish. In the irrelevant case the $W_i$ are all equal. The parameters used for the explicit case are: $\lambda = 3 \mu$ for relevant, $\lambda = 1 \mu$ for irrelevant and  $T=0.002 \mu$, $p/\mu=1$, $\kappa=1/\sqrt{2}$. For spontaneous: $\lambda=0.05 \mu$, $T=0.005\mu$, $p/\mu=1$, $\gamma=1.7$, the irrelevant case is simply Reissner-Nordstr\"om.
}%
\label{fig:profiles}
\end{figure}

Let us take a closer look at the possible classes of solutions one can get in our model \eqref{eq:action}. We first start from the case with purely explicit symmetry breaking \cite{Donos:2012js}. Without an explicit symmetry breaking source, the model always admits a translationally symmetric RN solution. When we turn on a small value of $\lambda$, the translations are broken by the field $B_2$, but the near horizon geometry of the black hole in the IR is not modified qualitatively. In particular, all the values of $W_i$ metric components are equal (see Fig.\,\ref{fig:profiles}). The deformation introduced by small $\lambda$ is thus \textit{irrelevant} and the profile of the $B_2$ field vanishes at the horizon. 
These solutions have been studied previously in \cite{Donos:2014oha}.

However, when we increase the strength of explicit TSB, a (quantum) phase transition takes place \cite{Donos:2012js}. At some critical value of $\lambda$ the TSB deformation becomes \textit{relevant} in the IR and modifies the near-horizon geometry qualitatively, one observes that the $B_2$ field is now finite at $u=1$ and the geometry becomes anisotropic: $W_1$ field diverges, while $W_2$ and $W_3$ fields vanish on the horizon, see Fig.\,\ref{fig:profiles} (left panel). In most of our examples, the parameters are 
\begin{equation}
\label{eq:explicit parameters}
\mbox{Explicit:} \qquad \kappa=1/\sqrt{2}, \ \  \gamma = 0, \ \ k/\mu=1, \ \ \lambda = 3 \mu
\end{equation}
and the phase transition happens at $\lambda \approx 2.7 \mu$.

This behavior can be understood \textit{analytically} by studying the exact $T=0$ near horizon asymptotes of the equations of motion. 
As it was shown in \cite{Donos:2012js} (as we also revisit in Appendix\,\ref{app:zero_T}), at zero temperature the near horizon 
behavior of the field profiles is ($r\equiv1-u$)%
\footnote{Note that in order to make connection to the results of \cite{Donos:2012js} one has to substitute $W_i = e^{2 v_i}$.}
\begin{align}
\label{eq:zeroT_explicit}
&\mbox{Explicit TSB:} & 
 W_1 &\sim W_1^0 r^{-2/3}, & W_2 &\sim W_2^0 r^{4/3}, & W_3 &\sim W_3^0 r^{2/3} \\
 \notag
&& A_t &\sim  A_t^0 r^{5/3}, & T &\sim  T^0 r, & & \\
\notag
&& B_2 &\sim w^0 + w^1 r^{4/3}, & A_2 &= 0, &  Q &= 0. 
\end{align}
These scalings match the qualitative behavior of our numerical solutions at small temperature on Fig.\,\ref{fig:profiles} and 
describe the new IR fixed point which is achieved once the system is deformed by a relevant explicit translation symmetry 
breaking potential. As we will show shortly, this fixed point is \textit{insulating}: the resistivity diverges at zero temperature.

The other possible class of solutions is driven by the \textit{spontaneous} translational symmetry breaking. This happens in absence of the explicit source $\lambda$ due to the dynamical instability induced by the $\gamma$ CS term in \eqref{eq:action}. At the critical temperature the field $A_2$ develops a nontrivial profile, which breaks translations spontaneously (see Fig.\,\ref{fig:bell}). As the temperature is lowered this profile backreacts on the geometry and drives the system to a new ground state. The helical pitch $k$ characterizing this spontaneous solution is fixed dynamically by the minimum of thermodynamic potential and depends on temperature as shown on Fig.\,\ref{fig:bell}. 
Unless otherwise stated, we concentrate on these thermodynamically stable configurations below. 
In what follows we will be mostly focusing on the case 
of $\gamma=3$ with $T_c \approx 0.223 \mu$ and $k_c\approx 2.18 \mu$.

Similarly to the relevant explicit case, the near horizon geometry in the new state is modified qualitatively. On the Fig.\,\ref{fig:profiles} (right panel) we see again that $W_1$ field diverges while $W_2$ and $W_3$ fields vanish. On top of that the off diagonal $t \omega_2$-component of the metric $Q(u)$ is now diverging at the horizon. Note that the spontaneous deformation is always relevant at the horizon, since it is driven by the instability of the homogeneous solution. 

Similarly to the relevant explicit case, we can access corresponding zero temperature groundstate analytically, as we discuss in more details in Appendix \ref{app:zero_T}. The near horizon asymptotes of the fields turn out to be very similar, except that the role of $B_2$ is now played by $A_2$ and there is a nonzero $Q$ component:
\begin{align}
\label{eq:zeroT_spont}
&\mbox{Spontanoeus TSB:} & 
 W_1 &\sim W_1^0 r^{-2/3}, & W_2 &\sim W_2^0 r^{4/3}, & W_3 &\sim W_3^0 r^{2/3} \\
\notag
&& A_t &\sim  A_t^0 r^{5/3}, & T &\sim  T^0 r, & & \\
\notag
&& B_2 &= 0, & A_2 &\sim b^0 + b^1 r^{4/3}, &  Q &\sim Q^0 r^{2/3}. 
\end{align}

In the purely spontaneous phase there is a Goldstone mode in the spectrum associated with homogeneous shifts along the axis of the helix. It witnesses the underlying translation symmetry of the action and will mediate the perfect conductance under applied electric field. In order to provide finite resistivity, we will introduce small source of explicit breaking $\lambda$, which will pin the Goldstone mode and move it to finite energy. As discussed in \cite{Andrade:2017cnc}, this will immediately promote the system to the insulating state, which is in the focus of our present study. 
It is worth mentioning that the Goldstone mode introduces a technical complication in our numerical schemes: as we have checked, we cannot reliably address solutions with very small $\lambda$ due to the fact that this light mode allows the numerical scheme to drift away from the true solution and renders in unstable. 
Therefore, throughout this work we will address small, but finite values of explicit pinning $\lambda$. 
As a rule, we use $\lambda$ which is much smaller then the spontaneous order parameter see Fig.\,\ref{fig:bell}, right. In most cases we will present the data for
\begin{equation}
\label{eq:pinned spontanoues parameters}
\mbox{Pinned spontaneous:} \qquad \lambda = 0.1 \mu, \ \ \gamma=3, \ \ \kappa=0, \ \  T_c \approx 0.223 \mu, \ \ k_c\approx 2.18 \mu.
\end{equation}

The extra $\lambda$ source requires a more careful treatment at zero temperature since it turns on the $B_2$ field in \eqref{eq:zeroT_spont}. We will consider the values of $\lambda$ which are small enough and do not drive the system in the relevant explicit TSB fixed point discussed above. However, even an irrelevant deformation of the $B_2$ field near horizon may play an important role, since in this case it is compared to zero,
e.g., it is a so called \textit{dangerously irrelevant} deformation. It can be analyzed using the zero temperature linearized equations of motion, see Appendix\,\ref{app:zero_T}. We find that it has the form\footnote{Note that the definition of $\alpha$ differs here from that of \cite{Donos:2012js}, where the analogous value is $2/3 + \delta$, $\delta=(\sqrt{1 + 120 \kappa^2} - 5)/6$}
\begin{equation}
\label{eq:lambda_deforamtion}
\delta B_2 \sim \delta B_2^0 r^\alpha, \quad \delta B_t = \delta B_2^0 \frac{12 \sqrt{3} \kappa}{\sqrt{1 + 120 \kappa^2} - 1} \frac{r^{\alpha + \frac{1}{3}}}{W_2^0} , \quad \alpha = \frac{1}{6}\left(\sqrt{1 + 120 \kappa^2} - 1 \right).
\end{equation}
Even though this mode will not play a crucial role in the subsequent discussion, we will show in the next Section why it is 
important to be aware of its existence.

\section{\label{sec:Inc_conductivity} Holographic conductivities}

The central object of the present study is the conductivity of the two ``insulating'' ground states described above. We will only focus on the conductivities along the helix director, since this is the direction in which the translational symmetry is broken. 
In the linear response approximation, we turn on an electric field ($E^x$) and temperature gradient ($\nabla^x T$) in the dual boundary theory and read off the induced expectation values for the electric current $J^x$ and heat current $Q^x$, obtaining therefore the full matrix of thermoelectric conductivities. 
In the bulk, these values correspond to the near boundary asymptotes of the gauge field $A_x$ and off-diagonal component of the metric $g_{tx}$, which one can write down as\footnote{Note that here we omit some logarithmic terms which are present due to the conformal anomaly, \cite{Henningson:1998gx, deHaro:2000vlm}.}
\begin{align}
\label{pert g A}
	\delta g_{tx} &= \delta g_{tx}^{(0)} + u^2 \delta g_{tx}^{(2)}  + u^4 \delta g_{tx}^{(4)} + \ldots \\
	\delta A_x &= \delta A^{(0)}_x + u^2 \delta A^{(2)}_x + \ldots
\end{align}
As it is discussed in \cite{Hartnoll:2009sz}, the leading branches of the solutions $\delta A^{(0)}_x$ and $\delta g_{tx}^{(0)}$ 
correspond to the sources, while the expectation values of dual operators $J^x$ and $Q^x$ are encoded in the subleading components 
$\delta A^{(2)}_x$ and $\delta g_{tx}^{(4)}$, see Eq.\,\eqref{Jx UV},\eqref{Qx UV}. The precise relation for the sources
reads
\begin{equation}
\label{eq:matrix_sigma}
\left( 	\begin{matrix}
 	\la J^x \ra \\ 
 	\la Q^x \ra
  	\end{matrix} \right) = \left( \begin{matrix}
 	\sigma & T \alpha \\ 
 	T \bar \alpha & T \bar \kappa
 	\end{matrix}  \right) \left ( \begin{array}{c}
 	i \omega (\delta A_x^{(0)} + \mu \delta g_{tx}^{(0)})  \\ 
 	i \omega \delta g_{tx}^{(0)} 
 	\end{array} \right ).
 \end{equation}


At the end of the day, in order to study the full conductivity matrix we solve the linearized equations of motion with the sources 
$\delta A^{(0)}_x$ and $\delta g_{tx}^{(0)}$ and read off the asymptotes of the solutions, the details are provided in Appendix\,\ref{app:AC_conductivities}. 
As discussed in \cite{Donos:2015bxe}, the presence of spontaneous magnetic currents, encoded in the $A_2$ field, introduces subtleties in the definition of the thermoelectric conductivities. However, these magnetic terms do not contribute to the conductivities along the helix director, as it trivially follows from their symmetry properties.

A central role in our study is played by a specific linear combination of the thermoelectric conductivities -- the so-called \textit{incoherent conductivity} $\sigma_{inc}$ introduced in \cite{Hartnoll:2007ih,Blake:2013bqa,Davison:2014lua,Davison:2015bea,Davison:2015taa} in the hydrodynamic approximation, addressed in spontaneous setups in \cite{Gouteraux:2018wfe,Amoretti:2017axe,Amoretti:2017frz} and further studied at finite frequency in \cite{Donos:2018kkm}. This quantity describes the propagation of the \textit{incoherent current} -- a combination of heat and electric currents, which has no overlap with momentum. As such, it is insensitive to the physics of momentum conservation/dissipation and $\sigma_{inc}$ is finite even in the absence of explicit TSB, when the momentum is conserved and all other conductivities diverge at zero frequency.  

The frequency dependent definition of \cite{Donos:2018kkm} yields
\begin{equation}
\label{sigma inc good}
		\sigma_{inc}(\omega) = \frac{1}{(\epsilon+p)^2}[ ( \epsilon + p - \mu \rho) ^2 \sigma(\omega) - 2 ( \epsilon + p - \mu \rho)  \rho T \alpha(\omega) + \rho^2 T \bar \kappa(\omega) ],
\end{equation}
\noindent where $\epsilon$ is the energy density, $p$ the pressure, $s$ is the entropy density, $T$ the temperature, $\rho$ the charge density, and $\sigma$, $\alpha$ and $\bar \kappa$  the components of the conductivity matrix \eqref{eq:matrix_sigma}. The overall normalization\footnote{We use a notation which slightly differs from \cite{Donos:2018kkm}, since $\sigma_{inc}|_{\mathrm{here}} = (\epsilon+p)^{-2}\sigma_{inc} |_{\mathrm{there}}$} 
was chosen so that 
$\sigma(\omega)|_{\omega \rar 0} = \sigma_{inc}(\omega) + \rho^2/(\epsilon+p) \cdot i \omega^{-1}$ in the case without explicit TSB. We will further discuss the features of $\sigma_{inc}$ in the finite frequency domain in Section\,\ref{sec:AC_conductivity}.

\section{\label{sec:DC_conductivity} DC response}

Let us first focus on the DC response. As we discuss in more detail in Appendix\,\ref{app:DC_formulas}, the DC conductivities can be calculated using the powerful formalism developed in \cite{Donos:2014cya,Donos:2015bxe,Banks:2015wha,Donos:2015gia,Donos:2017mhp,Donos:2018kkm}. It allows one to avoid solving the linearized equations and obtain the conductivities given only by the near horizon asymptotics 
of the background solutions. 

Using the expressions for the conductivities in terms of horizon data \eqref{equ:DC_formulae} we can follow their dependence with temperature substituting the numerical backgrounds of the two classes described in Sec.\ref{sec:model}. As far as we consider the backgrounds with finite source for explicit symmetry breaking $\lambda$, all the DC conductivity coefficients are finite and we can simply 
evaluate their combination \eqref{sigma inc good} and obtain the incoherent conductivity. 

\begin{figure}[t]
\centering
\includegraphics[width=1.0 \linewidth]{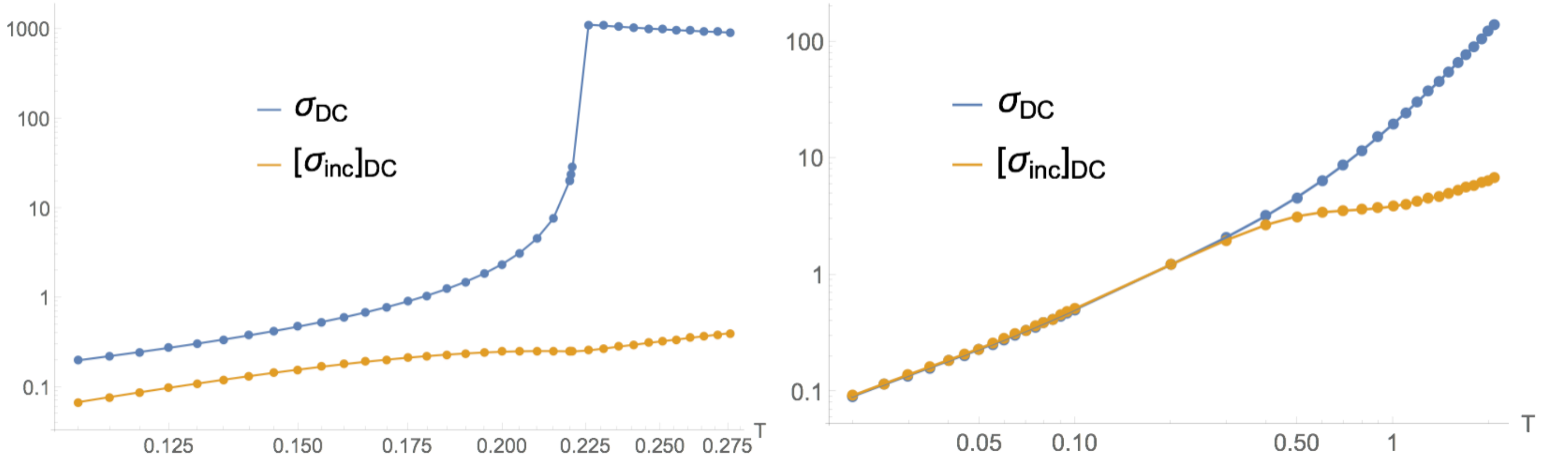}
\caption{Optical (blue) and incoherent (yellow) DC conductivities as a function of temperature for pinned spontaneous (left) and 
explicit (right) holographic insulators. For the pinned spontaneous case we take the thermodynamically preferred purely spontaneous 
solutions and place them on a lattice with $\lambda = 0.1\mu$ \eqref{eq:pinned spontanoues parameters}. In the explicit case we take $\lambda = 3\mu$, 
$k / \mu=1$ \eqref{eq:explicit parameters}. Both plots are shown in log-log scale. It is apparent that unlike in the explicit case, in the spontaneous case the incoherent conductivity only brings subleading contribution. }%
\label{fig:DC}
\end{figure}

We show our results for the temperature dependence of the DC electric and incoherent conductivities in Fig. 
\ref{fig:DC}. In the pinned spontaneous case (left panel) one can immediately see the steep drop of the conductivity at the phase transition temperature, where the spontaneous order is formed and gets immediately pinned. This metal-insulator transition we studied previously in \cite{Andrade:2017cnc,Mott} and will discuss it in more detail in Section\,\ref{sec:Omega}. In the purely explicit case (right panel) the phase transition is absent since there is no spontaneous order parameter. At low temperatures for both pinned spontaneous and relevant explicit cases, we observe that $\sigma_{DC}$ is an increasing function of $T$, showing the ``strange'' gapless insulating behavior.

Using \eqref{eq:zeroT_explicit} and \eqref{eq:zeroT_spont}, we are actually able to study this insulating regime analytically in the limit of small temperature. As we discuss in more detail in Appendix \,\ref{app:zero_T}, the horizon values for the fields in this case can be obtained from the zero temperature solution by evaluating it at finite distance from the horizon $r\rar r_h \sim T$. When we substitute these expressions 
into the horizon formula for DC electric conductivity \eqref{equ:DC_formulae} we obtain:
\begin{equation}
\label{eq:explicit_sig_scaling}
\mbox{Relevant explicit:} \qquad
\sigma_{DC} = \sqrt{\frac{W_2^0 W_3^0}{W_1^0}} \frac{(w^0)^2}{(w^0)^2 + (b^0)^2} T^{4/3} + O(T^{8/3}).
\end{equation}
The leading term is finite for the relevant explicit fixed point. Therefore we expect that the conductivity in this case will behave as $T^{4/3}$ at low temperature. The pinned spontaneous case is different: in this fixed point $w^0=0$ and the leading term vanishes, therefore we get
\begin{equation}
\label{eq:spont_sig_scaling}
\mbox{Pinned spontaneous:} \qquad
\sigma_{DC} = \frac{W_2^0}{(b^0)^2} \sqrt{\frac{W_2^0 W_3^0}{W_1^0}} T^{8/3} + O(T^{11/3})
\end{equation}
and we expect the scaling of the conductivity at low temperature to be $T^{8/3}$.
However, as we discussed earlier, when the leading term vanishes one has to pay a special attention to potentially dangerous irrelevant deformations introduced by small $\lambda$. Indeed, according to \eqref{eq:lambda_deforamtion} the pinning brings a small finite contribution $w_0 \sim T^{\alpha}$ and this produces an additional term in the conductivity of the order
\begin{equation}
\label{eq:delta_sigma_DC}
\delta_\lambda \sigma_{DC} = \delta \lambda \, T^{4/3 + 2 \alpha}, \quad \alpha = \frac{1}{6}\left(\sqrt{1 + 120 \kappa^2} - 1 \right).
\end{equation}
When $\alpha<2/3$, or $\kappa < 1/\sqrt{5}$, this term will define the \textit{leading} scaling of the conductivity at $T\rar 0$. In practice however, this dominant behavior can only be seen at extremely low temperatures, since it competes with the smallness of $\lambda$. As we show below, at reasonably finite $\lambda$ all our considered pinned spontaneous solutions tend to display scaling close to $8/3$, as expected from the leading order expression \eqref{eq:spont_sig_scaling}. 

\begin{figure}[t]
\centering
\begin{minipage}{0.49 \linewidth}
\includegraphics[width=1 \linewidth]{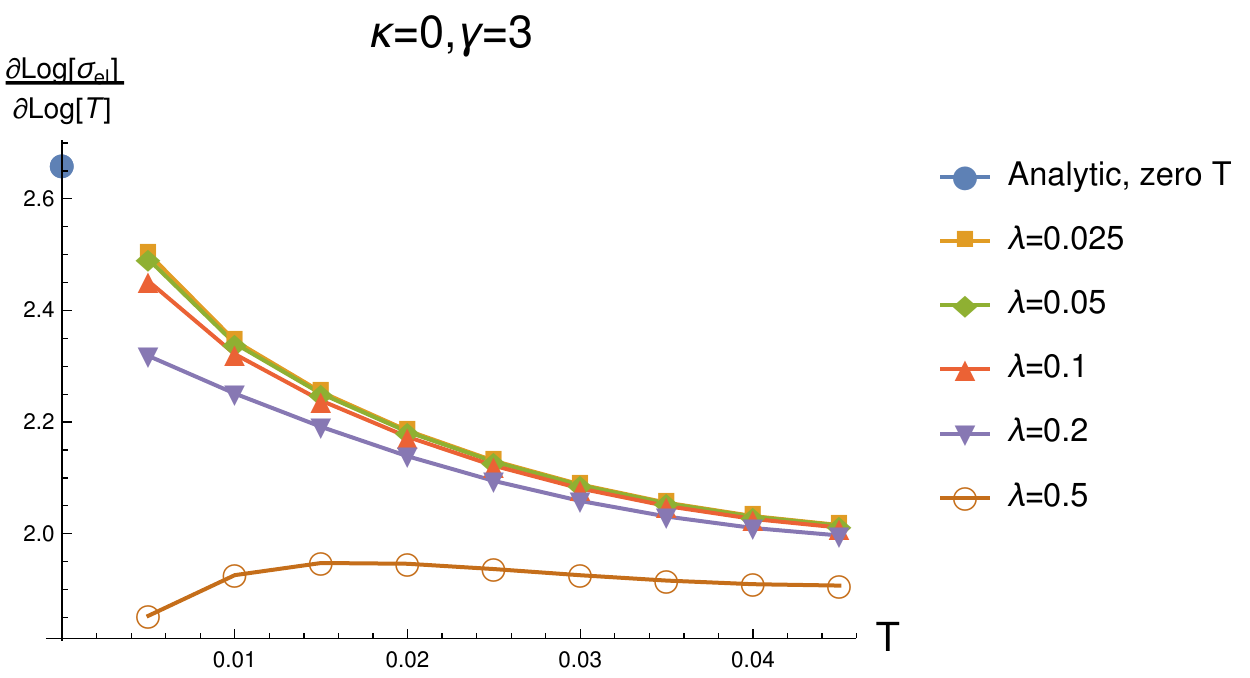}
\end{minipage}
\begin{minipage}{0.49 \linewidth}
\includegraphics[width=1 \linewidth]{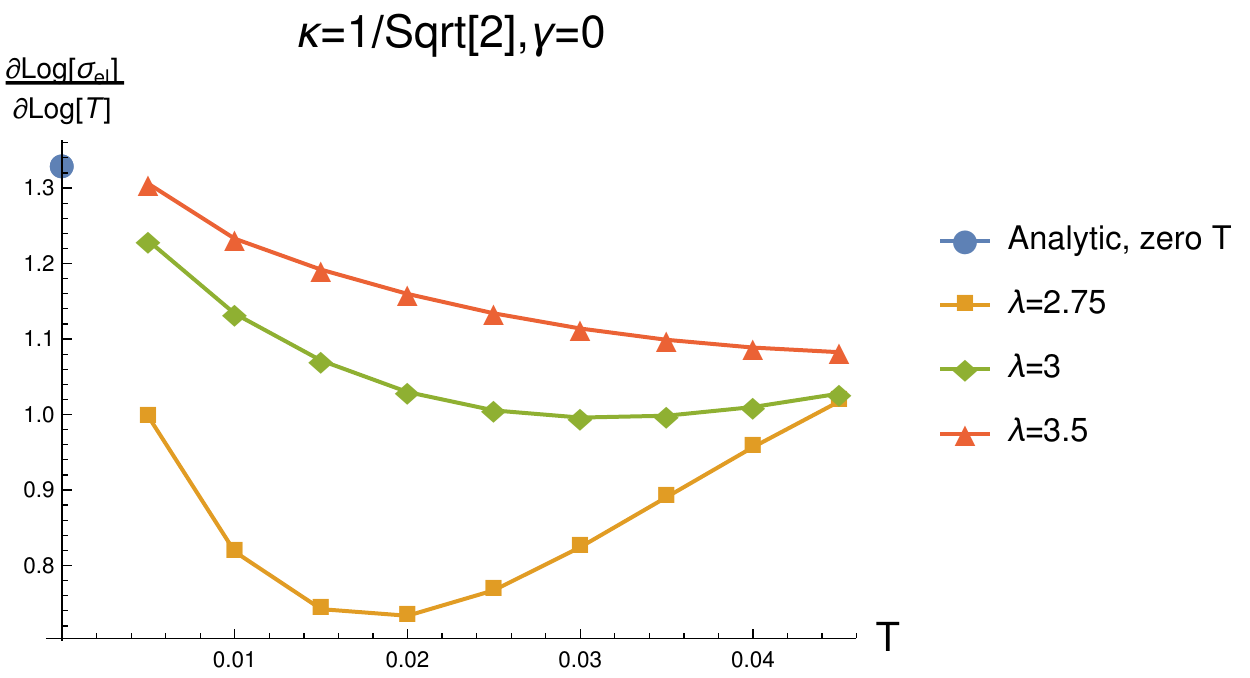}
\end{minipage}
\caption{
\textbf{Low temperature scaling of DC electric conductivity.} The logarithmic derivative shows how close the data is to a pure power law. \textit{Left panel:} Pinned spontaneous states with $p/\mu=1$, $\gamma=3$, $\kappa= 0$ and a range of $\lambda$. The states with small explicit source $\lambda$ clearly asymptote to an analytic value $8/3$, \eqref{eq:spont_sig_scaling} at low temperatures. As $\lambda$ grows the irrelevant deformation \eqref{eq:delta_sigma_DC} starts playing the leading role and the low temperature scaling changes.
\textit{Right panel:} Relevant explicit case with $p/\mu=1$, $\kappa=1/\sqrt{2}$ and a range of $\lambda$. All the states tend to the same asymptotic scaling $4/3$, predicted analytically at small temperature \eqref{eq:explicit_sig_scaling}.
}%
\label{fig:sigDCscalings}
\end{figure}

We can match these analytic predictions with our numerical data at small temperatures. In order to do so more precisely, 
we plot the logarithmic derivative of the DC electric conductivity versus temperature on Fig.\,\ref{fig:sigDCscalings}. For a pure power law the logarithmic derivative is a constant. On the plots, however, we see that its value changes in the region of temperatures which we consider, meaning that our numerical solutions are not cool enough to actually enter the zero temperature scaling limit. However we clearly see that the series we plot do extrapolate to the analytically computed values of $8/3$ for pinned spontaneous case and $4/3$ for relevant explicit case. We also observe that at larger $\lambda$ the scaling in the spontaneous case gets reduced due to the irrelevant deformation \eqref{eq:delta_sigma_DC}. This shows that the families of solutions which we consider do indeed correspond to the two distinct classes of IR fixed points, which we studied analytically at zero temperature and, importantly, these two fixed points yield very different scalings of the DC conductivity.

Another interesting difference between the two classes of holographic strange insulators becomes apparent when we compare the DC electric conductivity with DC incoherent conductivity, shown as yellow line on Fig.\,\ref{fig:DC}. On the plot for the relevant explicit case (right panel) we see that at low temperatures $T\lesssim 0.3 \mu$ the electric conductivity is completely determined by its incoherent part. This behavior is easy to understand: in this case the explicit translation symmetry breaking is relevant in IR and it gets stronger at lower temperatures. 
Therefore the momentum dissipation rate becomes large. This would lead to a strong suppression of the conductivity if it were 
completely governed by momentum conservation, but this does not apply to the incoherent part of the conductivity. The latter is insensitive to the translation symmetry breaking and therefore it dominates the transport in the regimes where the coherent contribution is suppressed. 

The situation is very different in the pinned spontaneous case, Fig.\,\ref{fig:DC} (left panel). Here we observe 
that $\sigma_{DC} \gg [\sigma_{inc}]_{DC}$ for all $T$. This is surprising since in this case one would expect the coherent contribution due to the Goldstone mode to be gapped by the pinning and therefore be moved to  finite frequencies. In this regard the situation would be similar to the explicit case: the coherent contribution is removed, therefore the only remaining part should be governed by the incoherent conductivity. 
However, as we can see, this logic does not apply since the incoherent conductivity only accounts for a small fraction of the total 
transport. Below we will uncover the origin of this mismatch, by studying the full frequency dependent profiles of the AC conductivity.

\section{\label{sec:AC_conductivity} AC conductivity}

In order to better understand the physics behind the two classes of hologrpahic strange insulators, 
we study the AC conductivity in the finite frequency domain. For this task we have to solve the linearized equations of motion for 
time-dependent perturbations as discussed in Sec.\,\ref{sec:Inc_conductivity}, see also Appendix\,\ref{app:AC_conductivities}. 
We will also get additional information by studying the  quasinormal modes, which describe the spectrum of linearized 
perturbations, see Appendix\,\ref{app:AC_conductivities}. The formula for the incoherent conductivity 
\eqref{sigma inc good} allows us to evaluate it at finite frequency, which will be crucial for our studies. 
It should be noted that in \cite{Donos:2018kkm} this formula was derived with the assumption that 
translational symmetry is not broken explicitly. Therefore, it is not exactly applicable in the cases with finite 
explicit TSB source $\lambda$ which we study. 
Nonetheless, as we will see below, \eqref{sigma inc good} gives very good results at small $\lambda$ 
and works reasonably well in the case of the explicit relevant insulator with larger $\lambda$.

\begin{figure}[t]
\centering
\includegraphics[width=1 \linewidth]{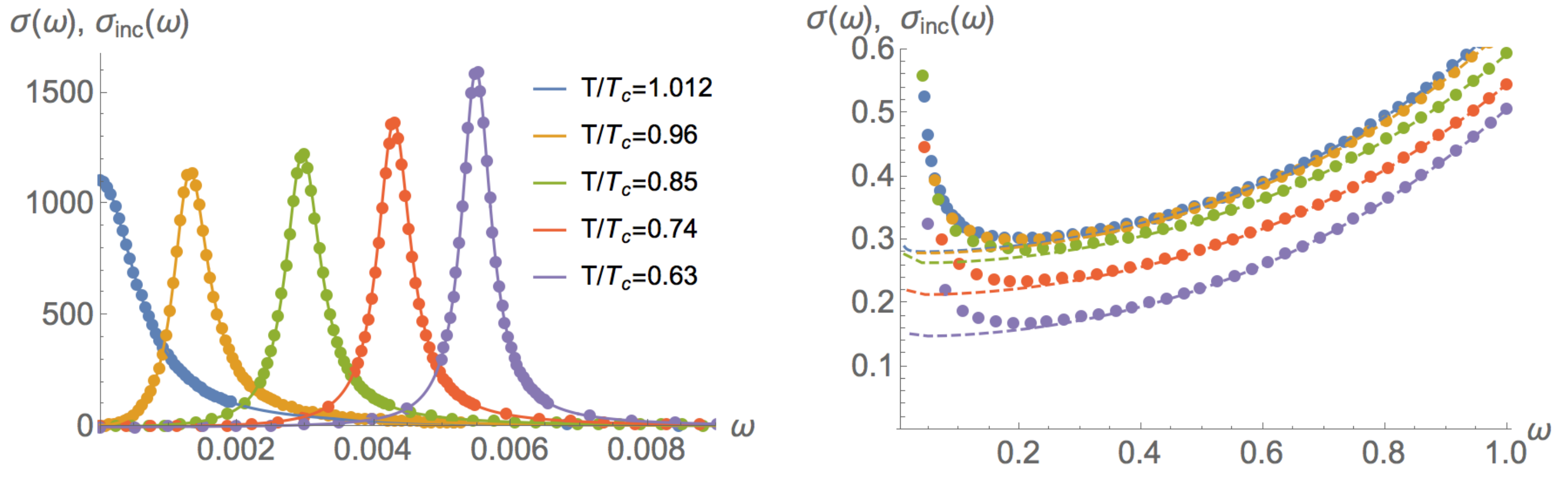}
\caption{AC optical and incoherent conductivities for the pinned spontaneous insulator \eqref{eq:pinned spontanoues parameters} at selected values of the temperature. We show the optical values with dots, and the incoherent with a dashed line.
The left panel zooms into the low frequency range, while the right panel show the high frequency results, one can see the right shoulders of the coherent peaks shown on the left panel, 
which arise in the low frequency region in the optical conductivity.}%
\label{fig:AC}
\end{figure}

We have numerically extracted the optical and incoherent conductivities as explained in Sec.\,\ref{sec:Inc_conductivity}, see Fig.\,\ref{fig:AC} for the results. 
We have checked that the small frequency limit is in excellent agreement with the
DC values obtained from the horizon formulae, see Fig.\,\ref{fig:DeltaSigmaDC} in App.\,\ref{app:DC_formulas}
for the exact comparison.

\begin{figure}[t]
\centering
\includegraphics[width=1 \linewidth]{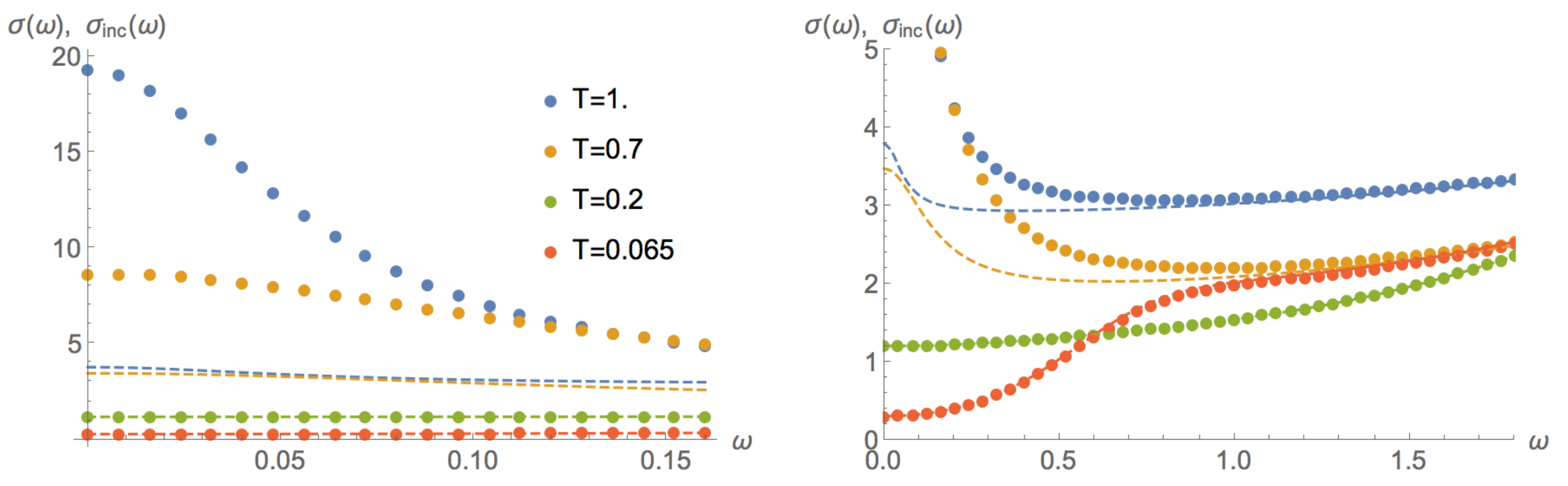}
\caption{AC optical and incoherent conductivities for the explicit insulator \eqref{eq:explicit parameters} at selected
values of the temperature. We show the optical conductivity with a solid line, and the incoherent one with a dashed line.
The left panel zooms into the low frequency range, while the right panel show the high frequency results.}%
\label{fig:AC_exp}
\end{figure}

Let us first focus on the explicit relevant case. At high temperature, the relevant TSB has the weakest effect and a 
sharp peak is seen on the AC conductivity plot shown on Fig.\,\ref{fig:AC}.
The peak corresponds to a momentum-mediated transport, and it has small width due to weak momentum dissipation. 
The dashed line on the same plot corresponds to the incoherent AC conductivity. Its remarkable feature is that it coincides with electric conductivity everywhere at higher frequencies, but precisely ``cuts off'' the pole at low frequency. This behavior follows from 
the very definition of incoherent transport: it is orthogonal to momentum therefore it completely neglects 
the momentum contribution.
We will analyze this in more detail in the next Section, where we will see that the difference between 
the optical and incoherent contributions indeed corresponds to a Drude peak at hight temperatures. 

%

As the temperature is lowered, the TSB becomes stronger, momentum dissipates faster and the peak gets broader. 
At some point at low temperature one cannot discern the peak shape anymore and the electric conductivity coincides with the incoherent one everywhere in the frequency domain. This corresponds to the temperature $T \sim 0.3$ as seen on the DC 
conductivity plot, Fig.\,\ref{fig:DC}, where the electric conductivity gets completely accounted for by the incoherent part.

\begin{figure}[t]
\centering
\includegraphics[width=1 \linewidth]{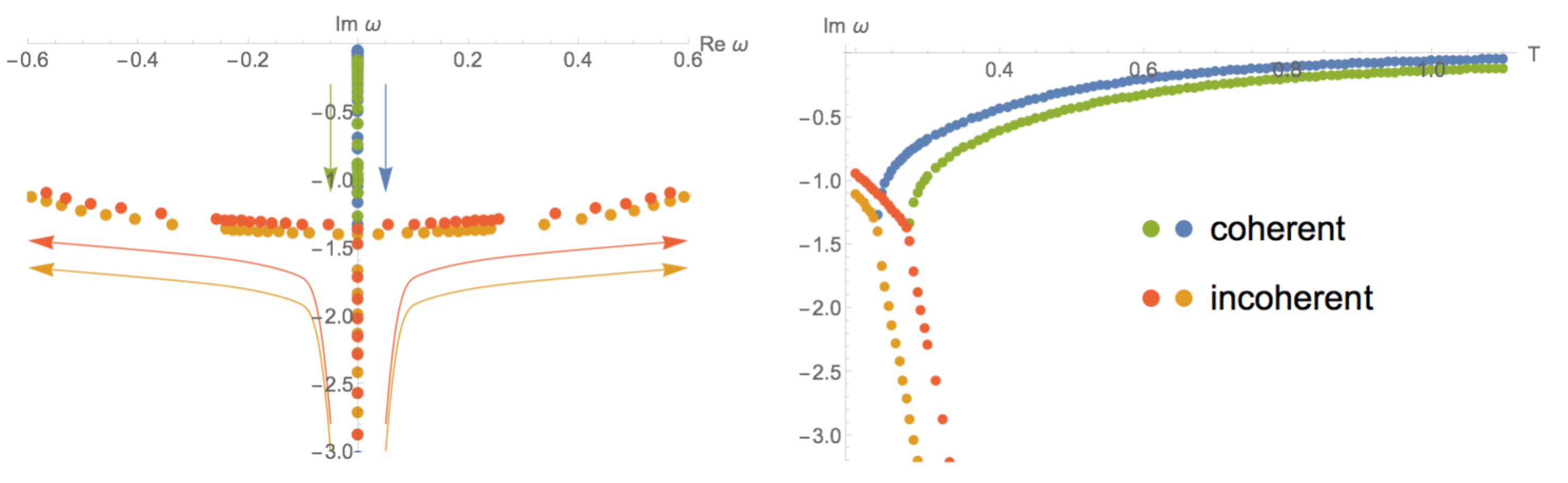}
\caption{Four lowest QNMs for the explicit case \eqref{eq:explicit parameters} with varying temperature. The green and blue dots show the  coherent modes corresponding to (approximate) momentum conservation. These would be the two sound modes in the translationally invariant case, which turn into two diffusive modes in presence of momentum relaxation. The red and yellow dots are incoherent, non-hydrodynamic modes. The left panel shows the motion of 
these modes in the complex plane, the arrows pointing towards lower temperatures. 
We observe two collisions on the imaginary axis as the temperature is lowered.
After the collision the modes acquire a non-zero real part and the coherent modes can no longer be identified. We show the imaginary parts of these modes as a function of 
temperature on the right panel.}%
\label{fig:QNM explicit}
\end{figure}

\begin{figure}[ht]
\centering
\includegraphics[width=1 \linewidth]{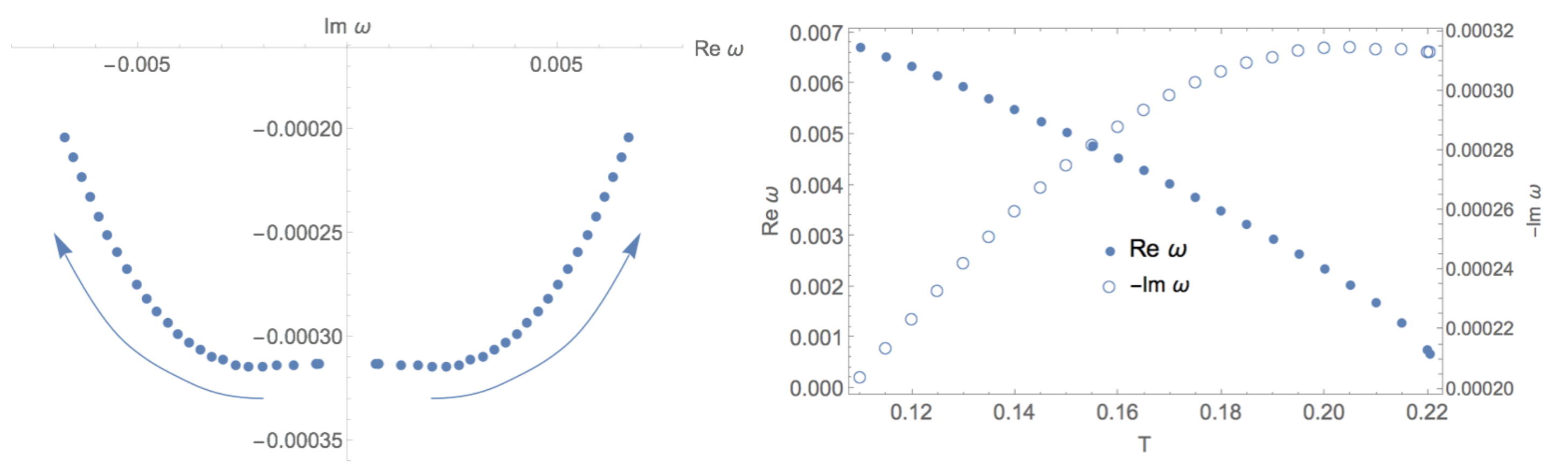}
\caption{First QNM for the pinned spontaneous case \eqref{eq:pinned spontanoues parameters} with varying temperature. The left panel shows the motion of the QNM in the complex
plane, the arrow pointing towards lower temperatures as $T/T_c$ changes from 1 to 0.63, same range as Fig.\ref{fig:AC}. The right panel shows both real and imaginary parts as a function 
of temperature, with the corresponding scales given on the left and right vertical axes. As the temperature is lowered, the QNM comes closer to the real axis and the coherent contribution to the transport never disappears.}%
\label{fig:QNM_Mott_vsT}
\end{figure}

This transition can be very well understood by looking at the behavior of QNMs, shown on Fig.\,\ref{fig:QNM explicit}. At high temperature one observes an isolated purely imaginary quasinormal mode very close to the real axis. This corresponds to a weakly dissipating momentum and gives rise to a sharp peak in the AC conductivity plot.  These low lying modes can be accounted very well by the hydrodynamics \cite{Policastro:2002se,Kovtun:2004de}. As long as momentum is an (almost) conserved quantum number, one can treat them 
hydrodynamically and obtain the late time dynamics of the system, corresponding to the quasinormal mode with small imaginary part. 
Importantly, there are other quasinormal modes in the spectrum, which cannot be accounted for by hydrodynamics 
and which lie further from the real axis (shown in red/yellow on Fig.\,\ref{fig:QNM explicit}). These are ubiquitous in holography and have been studied previously at length \cite{Cardoso:2001bb,Berti:2003ud,Michalogiorgakis:2006jc,Balatsky:2014lsa}. Importantly for us, since they are not accounted for by hydrodynamics, they do not correspond to the physics of momentum conservation and therefore, collectively, describe the incoherent part of transport. 

As the temperature is lowered, the momentum dissipation mode moves further down in the complex $\omega$ plane, 
 and consequently the peak getting broader. At some point it gets so deep into the lower imaginary half plain that it meets with one of the lower lying ``incoherent'' QNMs and recombines, as seen on Fig.\,\ref{fig:QNM explicit}. At this temperature, the hydrodynamic mode and the corresponding peak can no longer be identified, i.e., transport becomes fully incoherent. This constitutes the first main result of our study: \textit{the holographic insulating state arising from the relevant explicit translation symmetry breaking is fully characterized by the incoherent conductivity.} 

Next we turn to the pinned spontaneous state. Its AC conductivity is shown on the left panel of Fig.\,\ref{fig:AC}. One can start, again, at high temperature, when the sharp Drude peak is seen. It is much sharper then that of the explicit relevant case since the value of $\lambda$, which controls its width, is much larger in the latter. As we lower the temperature we hit the phase transition, where the spontaneous helical structure is formed in the profile of field $A_2$. In the pure case without explicit symmetry breaking, the arising gapless Goldstone mode would account for the full spectral weight corresponding to momentum-mediated transport -- the delta function in AC response. 
In the pinned case, however, the Goldstone mode gets immediately gapped and drives all this spectral weight to the finite ``pinning'' frequency $\omega_0$. This mechanism was discussed in detail in \cite{Delacretaz:2017zxd,Delacretaz:2016ivq,Andrade:2017cnc} and it is responsible for a dramatic drop of DC conductivity at critical temperature. 
As the temperature is further lowered, the pinning frequency grows and the peak corresponding to the momentum-mediated, coherent, transport is moved to higher energies. 
Looking at the profile of the incoherent conductivity we see that it accurately removes the peak and at the same time matches well with the higher frequency incoherent tails of the plot. This matching at high frequencies provides a nontrivial check to our treatment of the incoherent conductivity. Note also that, in agreement with the horizon calculation of Sec.\,\ref{sec:DC_conductivity}, the zero frequency limits of electric and incoherent conductivities differ by an order of magnitude. 

We can again obtain more insight by studying the quasinormal modes. The results for pinned spontaneous case are shown on Fig.\,\ref{fig:QNM_Mott_vsT}. We see that at the critical temperature the hydrodynamic quasinormal mode, responsible for the Drude peak, acquires a real part, the pinning frequency, and moves off the imaginary axis. Contrary to the relevant explicit case, discussed above, as the temperature is further lowered the mode gets closer to the real axis and is never dissolved in the swarm of lower lying incoherent QNMs. This shows 
the crucial difference between the pinned spontaneous and relevant explicit hologrpahic insulators: \textit{in the pinned spontaneous 
case, the coherent momentum-mediated transport is always seen in the spectrum}. As we will see below this is the reason why the incoherent conductivity does not fully account for the transport in this case.

\section{\label{sec:Omega} Coherent transport}

\begin{figure}[th]
\centering
\includegraphics[width= 0.9 \linewidth]{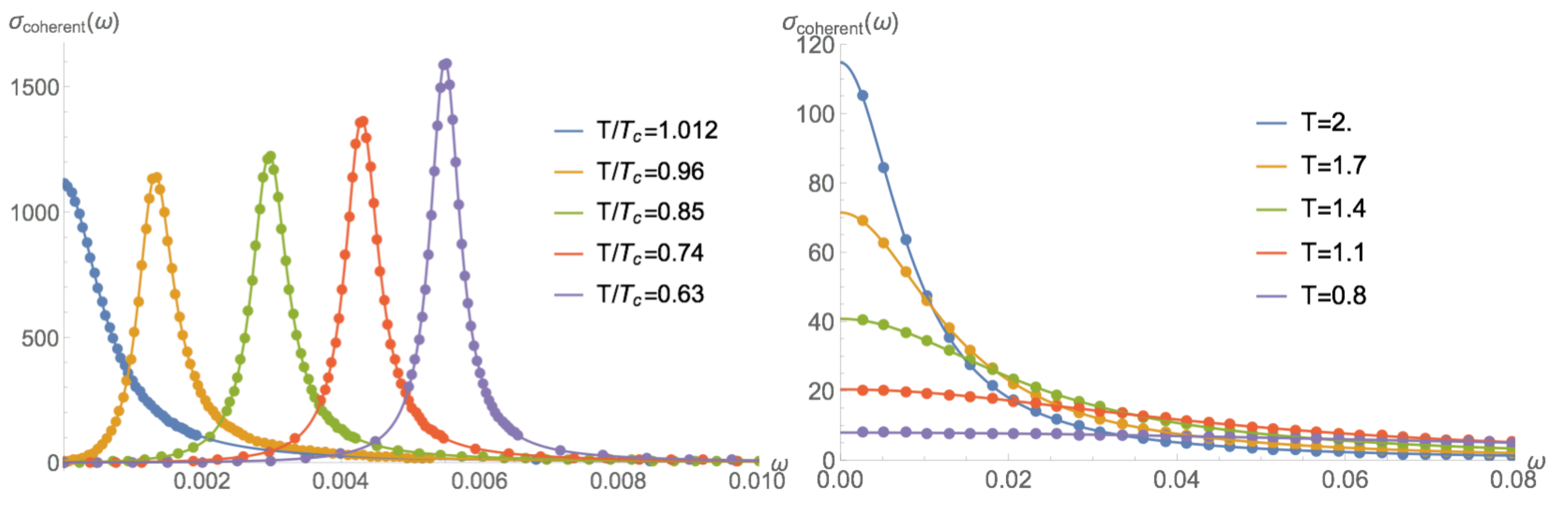}
\caption{Coherent conductivity for the pinned spontaneous \eqref{eq:pinned spontanoues parameters} (left) and explicit \eqref{eq:explicit parameters} (right) cases as a function 
of frequency. We show the numerical results as data points and corresponding fits to the models \eqref{drude}, \eqref{AC model}  with solid lines.}%
\label{fig:scoher}
\end{figure}

As we explained above, we have identified some situations in which transport is mediated by a hydrodynamic 
mode which is excluded in the incoherent conductivity. In this section we will focus specifically on this
contribution, therefore it will be convenient to consider the difference 
\begin{equation}
\label{eq:coherent}
	\sigma_{coherent}(\omega):= \sigma(\omega) - \sigma_{inc} (\omega).
\end{equation}
Importantly, as we noticed in the previous Section, $\sigma_{inc}(\omega)$ fits the high frequency tails of the electric 
conductivity very well, so the {\it coherent conductivity} defined via \eqref{eq:coherent}, has an isolated peak as its only 
feature and vanishes at larger $\omega$. We show the data for the coherent conductivity in Fig.\,\ref{fig:scoher}. 

\begin{figure}[th]
\centering
\includegraphics[width= 0.4 \linewidth]{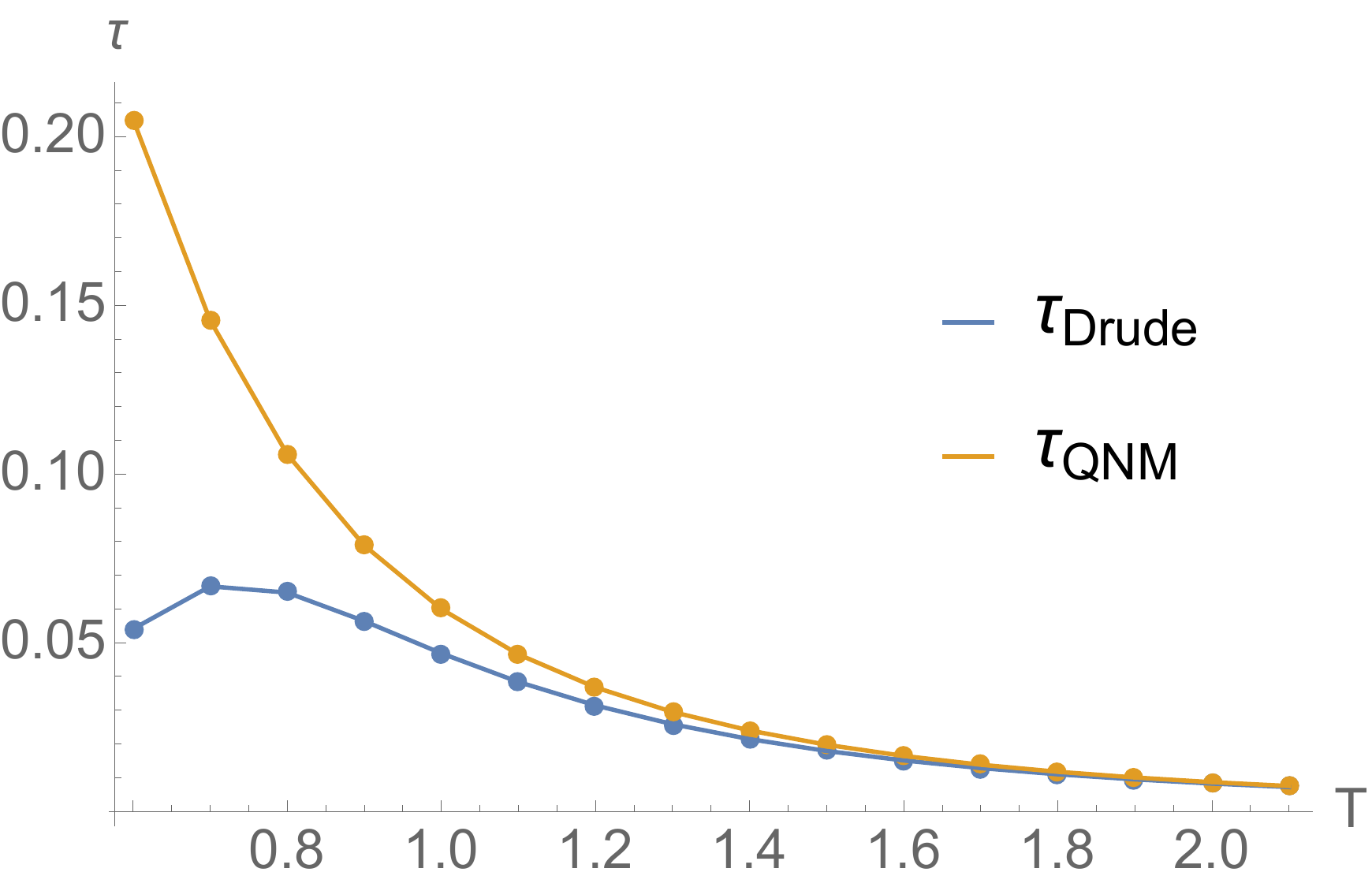}
\caption{Characteristic times extracted from a fit of the coherent conductivity to the Drude formula \eqref{drude} (blue), 
and from the imaginary part of the lowest lying QNM (yellow), as a function of temperature for the explicit insulator \eqref{eq:explicit parameters}.}%
\label{fig:tau_explicit}
\end{figure}

In the case of the explicit insulator, we expect transport to be coherent at high temperatures, where
there is a mode that lies near the origin which is well separated from the rest of the excitations.
We observe that the coherent conductivity can be adequately fitted to the Drude formula 
\begin{equation}
\label{drude}
	\sigma_{Drude}(\omega) = \frac{\sigma_{DC}}{1 - i \omega \tau}
\end{equation}
\noindent at hight temperatures. Indeed, extracting $\tau$ by fitting to the AC coherent conductivity,
we find that there is good agreement with the characteristic time given by the imaginary part of the lowest
lying QNM. We show this comparison in Fig.\,\ref{fig:tau_explicit}.
Upon lowering the temperature, the peak becomes broader and approximation to the Drude formula worsens,  
until we reach the fully incoherent regime. 

Let us now turn to the pinned spontaneous holographic insulator, where the coherent contribution largely dominates at all temperatures. 
The physics of this metal-insulator transition, which happens upon condensation of the spontaneous TSB order parameter can be well understood as a Goldstone mode acquiring a gap. This mechanism is captured by hydrodynamic approximation, as it has been shown recently on \cite{Delacretaz:2017zxd,Delacretaz:2016ivq}. It is actually expected from our study above: the corresponding quasinormal mode always lyes close to the real axis and therefore should be treatable by hydrodynamics.

Based on the treatment of \cite{Delacretaz:2017zxd,Delacretaz:2016ivq} we can introduce the phenomenological model for the electric conductivity in the pinned Goldstone case, whose coherent part is simply
\begin{equation}
\label{AC model}
\sigma_{coherent}(\omega) = A_0 \frac{\tilde{\Omega}- i \omega}{(\Gamma - i \omega)(\tilde{\Omega} - i \omega) + \omega_0^2}, \qquad A_0 = \frac{\rho^2}{\mu \rho + s T}.
\end{equation}
Here $A_0 = \rho^2/(\mu \rho + s T)$ equals the static momentum susceptibility $\chi_{PP}$, $\omega_0$ is the pinning frequency -- the gap introduced in the spectrum of pseudo-Goldstone by the explicit TSB source, $\Gamma$ -- the momentum relaxation rate. The $\tilde{\Omega}$ parameter has been associated in \cite{Delacretaz:2017zxd} with the phase relaxation rate (denoted $\Omega$), which can be included in the Josephson relation for the Goldstone mode. We intentionally use a tilde for this coefficient since its precise physical interpretation will not be important for us: we only use \eqref{AC model} as a phenomenological model to fit our results. 
Moreover, as we will discuss below, the phase relaxation rate may not be the dominant physical effect setting the value of $\tilde{\Omega}$.

It has already been shown in our previous work \cite{Andrade:2017cnc}, that the model \eqref{AC model} describes the AC conductivity across the spontaneous phase transition very well. It was found in particular that the parameters near the phase transition behave as $\omega_0\sim \lambda \la J \ra$ and $\Gamma \sim \lambda^2$, where $\la J \ra$ is the spontaneous helical current -- the order parameter of TSB, see Fig.\,\ref{fig:bell}. Here we extend this analysis to lower temperatures and focus in particular on the contribution to the DC conductivity, and the role of the parameter $\tilde{\Omega}$, which was not considered in \cite{Andrade:2017cnc}.

\begin{figure}[th]
\centering
\includegraphics[width=0.8 \linewidth]{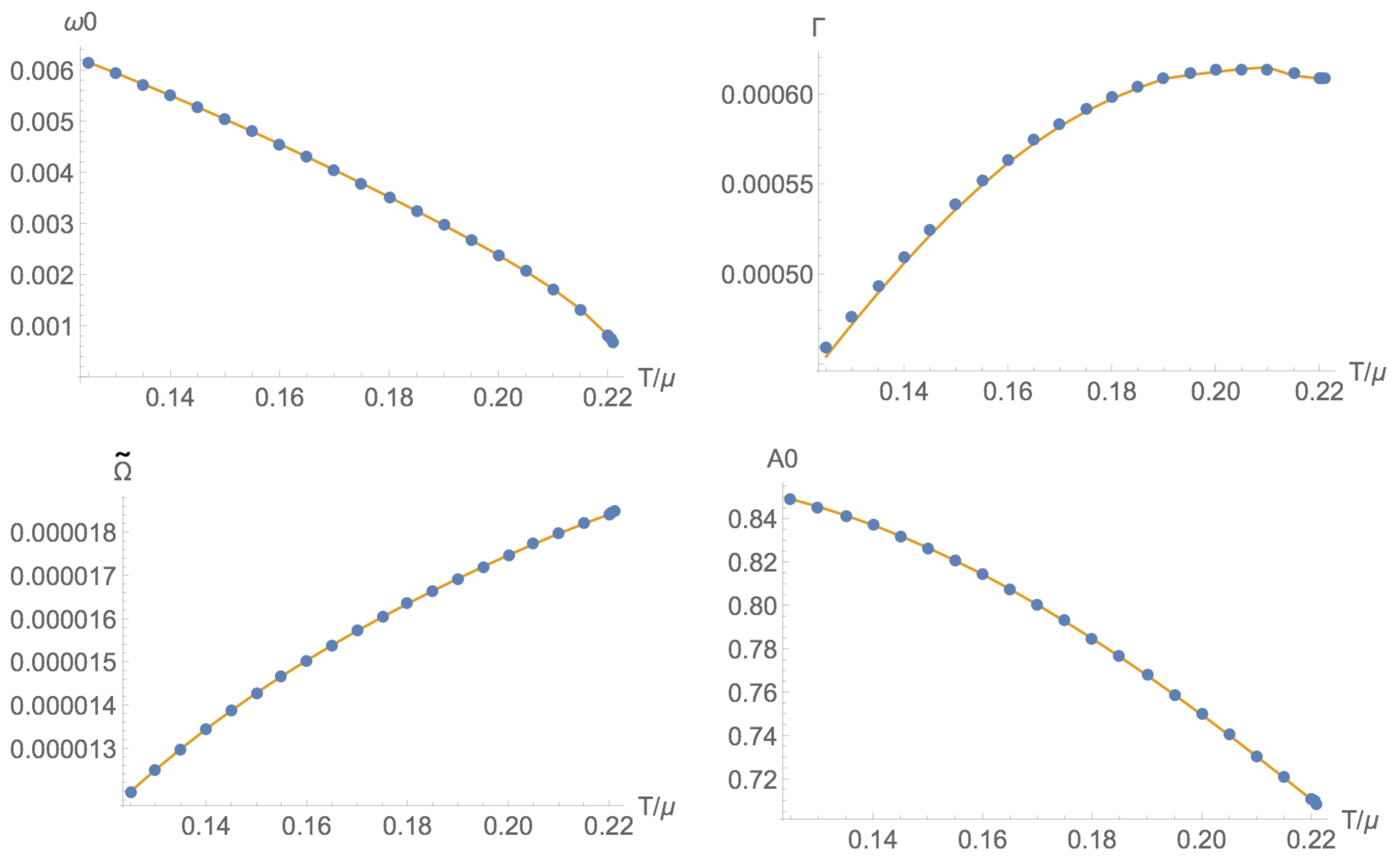}
\caption{Fitting parameters of the model \eqref{AC model} for pinned CDW, describing the pinned spontaneous data \eqref{eq:pinned spontanoues parameters} as a function of $T$. The data points denote the fit parameters extracted from $\sigma(\omega)$, while the solid lines represent different cross-checks. In the case of $\Gamma$, $\omega_0$
and $\tilde{\Omega}$ these are the computation of the QNMs combined with the extraction of $\sigma(- i w)$. The solid line in $A_0$ was obtained from the thermodynamical quantities on the right hand side of \eqref{AC model}.}%
\label{fig:fit_Mott_vsT}
\end{figure}

Since we have access to the full frequency dependent incoherent conductivity, we can exclude it from the data and extract the parameters of the model \eqref{AC model} very precisely by fitting the coherent part of the conductivity \eqref{eq:coherent}. 
We find that the model describes our AC data very well (see Fig.\,\ref{fig:scoher}), see Fig.\,\ref{fig:fit_Mott_vsT} for the dependence
of the fit parameters as a function of temperature.
We can also extract $\omega_0$ and $\Gamma$ from the position of the leading quasinormal mode (see Fig.\,\ref{fig:QNM_Mott_vsT}) and these are in agreement with ones obtained from the AC fit up to less than $1 \%$.
In addition, we have computed $\sigma(- i w)$ for real $w$, which allows us to easily extract
$\tilde{\Omega}$ as $\sigma(- i \tilde{\Omega})_{\mathrm{coherent}} =0$. Again, we find better than $1\%$ agreement between this calculation and the fits for $\sigma(\omega)$. Moreover, we have checked that the Onsager relation at finite frequency $\alpha(\omega) = \bar \alpha(\omega)$ is satisfied within good accuracy.

The result of all this precise analysis is the surprising finding that although numerically quite small, the parameter $\tilde{\Omega}$ is clearly nonzero for all the data sets which we have analyzed. One would be surprised by that since our ansatz \eqref{ansatz} clearly cannot take into account the dynamical topological defects, responsible for the phase relaxation rate as suggested in \cite{Delacretaz:2017zxd}. However, our data shows unambiguously that $\tilde{\Omega}$ is indeed. We see this not only in the AC conductivity fits, but also in the DC conductivity plots, as we discuss below.

We have seen in the previous Section that the incoherent DC conductivity in the pinned spontaneous
case is orders of magnitude smaller than the electric DC value. Given our discussion above, it is clear that the coherent part is responsible for this large difference. 
More concretely, note that at zero frequency the ansatz \eqref{AC model} leads to 
\begin{equation}
\label{sigma inc DC}
	\sigma_{DC}= [\sigma_{inc}]_{DC} +  A_0 \frac{\tilde{\Omega}}{\Gamma \tilde{\Omega} + \omega_0^2} 
\end{equation}
We see here that in order to get $\sigma_{DC} \gg [\sigma_{inc}]_{DC} $, it is crucial to have
$\tilde{\Omega} \neq 0$. 
This large coherent contribution might seem surprising since in the data shown in Fig.\,\ref{fig:fit_Mott_vsT}, we observe that $\tilde{\Omega}$
is one order of magnitude smaller than $\Gamma$ and two orders of magnitude smaller than $\omega_0$. 
However, considering the structure of \eqref{sigma inc DC} in more detail, we note that 
the difference $(\sigma_{DC}- [\sigma_{inc}]_{DC})$ is controlled by the ratio $\tilde{\Omega}/\omega_0^2$, which is of order 1.

\begin{figure}[th]
\centering
\includegraphics[width=0.6 \linewidth]{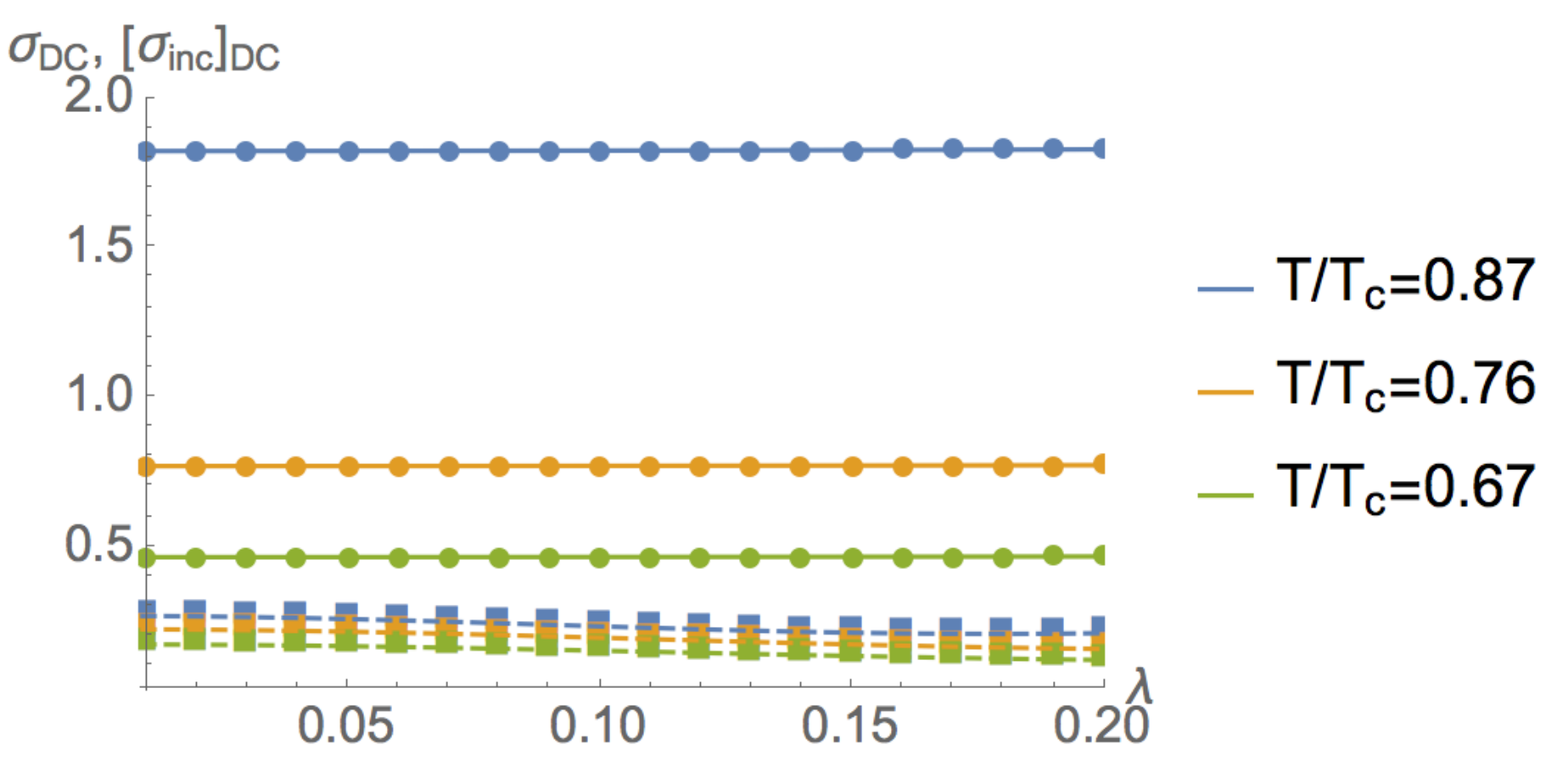}
\caption{DC conductivity (circles) and incoherent conductivity (squares) in the pinned spontaneous case \eqref{eq:pinned spontanoues parameters} as a function of $\lambda$
for selected temperatures. We join the data points with a solid and dashed line, respectively, to guide the eye.}%
\label{fig:DC_vs_lambda}
\end{figure}

Because of the importance of this effect, and given the large scale separation between the relevant physical parameters, we have performed a more detailed analysis of the parameter $\tilde{\Omega}$. 
It is very instructive to uncover the effect of the explicit symmetry breaking on the full electric conductivity of the pinned spontaneous insulator and on $\tilde{\Omega}$ in particular. 
For this purpose we studied a series of solutions as a function of $\lambda$ at fixed temperature, keeping therefore the spontaneous order parameter the same. 
The behavior of the conductivities as a function of $\lambda$ is shown on Fig.\ref{fig:DC_vs_lambda}. As expected on general grounds,  the incoherent conductivity is largely insensitive to the strength of explicit symmetry breaking, which controls the momentum relaxation rate. However it is quite surprising that the full electric conductivity, which is mostly controlled by the momentum sensitive coherent part, behaves in a similar way. Firstly, one expects the electric conductivity to diverge at $\lambda = 0$, since in this limit momentum is conserved and the transport is ballistic. Secondly, considering \eqref{sigma inc DC}, taking into account that $\Gamma \sim \lambda^2$ and $\omega_0 \sim \lambda$ \cite{Andrade:2017cnc}, and assuming constant $\tilde{\Omega}$, one would expect $\sigma_{DC}$ to diverge as $1/\lambda^2$. However this is not the case: $\sigma_{DC}$ behaves as a constant at small $\lambda$.

\begin{figure}[th]
\centering
\includegraphics[width=0.8 \linewidth]{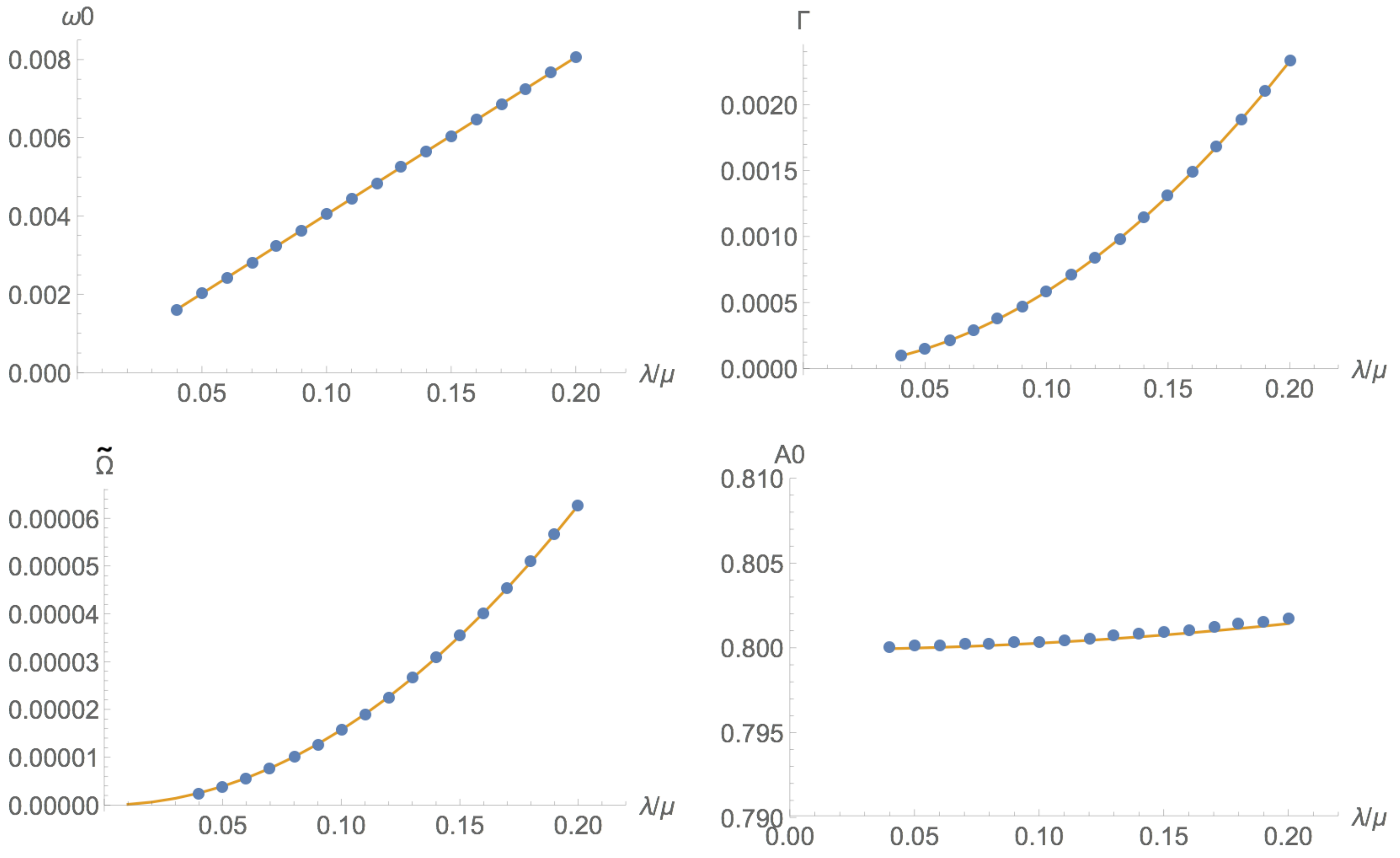}
\caption{
Fitting parameters of the model \eqref{AC model} for pinned CDW, describing the pinned spontaneous data \eqref{eq:pinned spontanoues parameters} as function of $\lambda$. The data points denote the fit parameters extracted from 
$\sigma(\omega)$, while the solid lines represent different cross-checks. In the case of $\Gamma$, $\omega_0$
and $\Omega$ these are the computation of the QNMs combined with the extraction of $\sigma(- i w)$. The solid 
line in $A_0$ was obtained from the thermodynamical quantities on the right hand side of \eqref{AC model}.}%
\label{fig:fit_Mott_vsLambda}
\end{figure}

\begin{figure}[th]
\centering
\includegraphics[width=0.5 \linewidth]{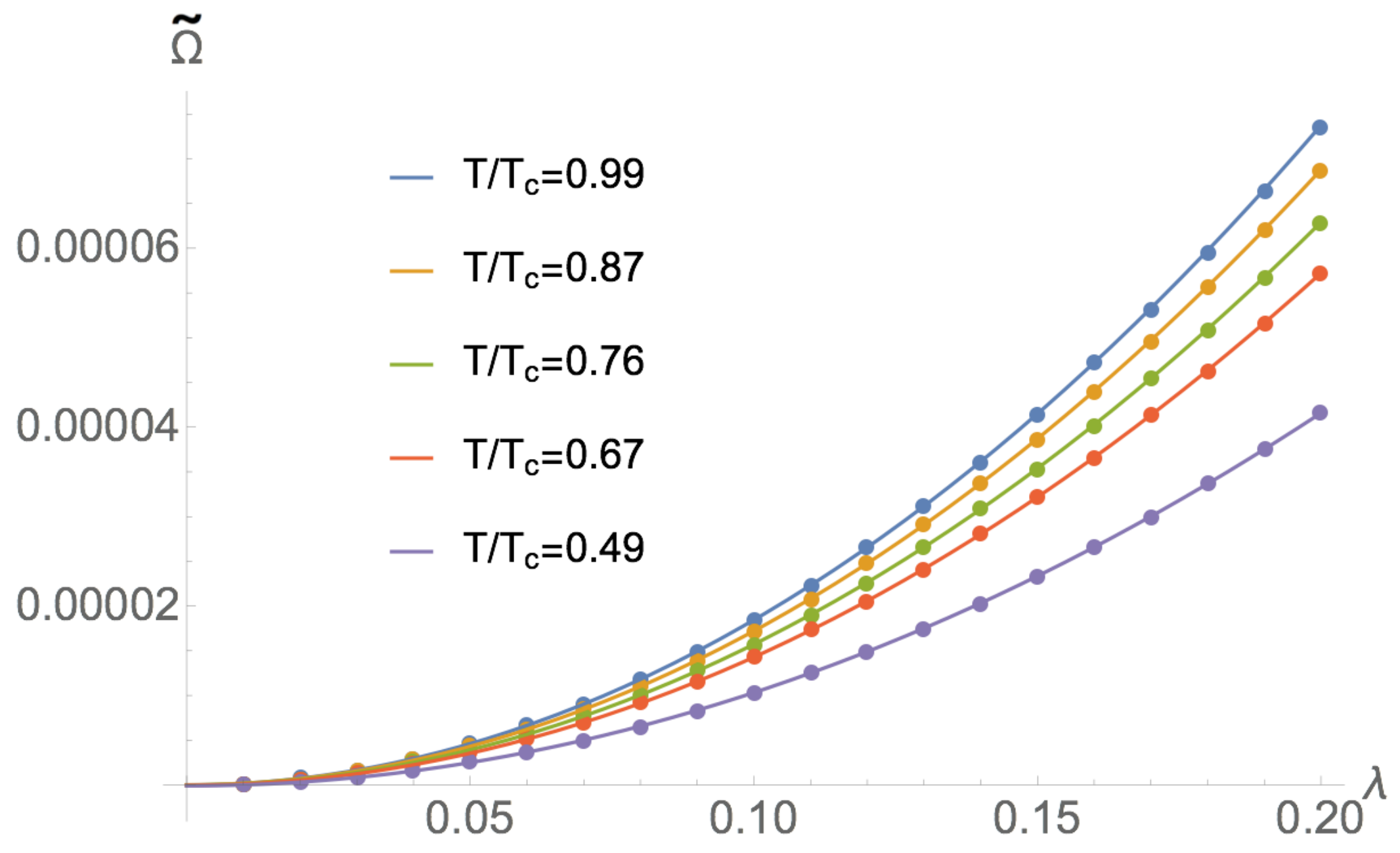}
\caption{Phase relaxation $\Omega$ as a function of the explicit breaking parameter $\lambda$.
The solid lines are fits to the data of the form $\Omega = a + b \lambda^c$, for which we obtain 
$a \sim 10^{-8}$, $b \in (1 \cdot 10^{-3}, 1.8 \cdot 10^{-3})$, $c = 2 \pm 10^{-4}$}%
\label{fig:OmegavsLambda}
\end{figure}

The more precise analysis of the DC and AC data at different $\lambda$, presented on Fig.\,\ref{fig:fit_Mott_vsLambda} points towards 
the resolution. One clearly discerns that the fitting parameter $\tilde{\Omega}$ is proportional to $\lambda^2$. This behavior is robust and is observed at all temperatures as shown on Fig.\,\ref{fig:OmegavsLambda}. This finding is the second major result of our study: \textit{The resistivity of the pinned spontaneous holographic strange insulator is mostly controlled by coherent contribution, which is nonzero due to finite parameter $\tilde{\Omega}$ in \eqref{AC model}.} The latter, however, is of the different physical origin then the phase relaxation rate due to topological defects considered in \cite{Delacretaz:2017zxd,Delacretaz:2016ivq}. It vanishes at $\lambda = 0$ and therefore cannot be attributed to the features of the pure spontaneous superstructure, as it would be the case for the moving topological defects, suggested in \cite{Delacretaz:2017zxd}.

One possible explanation of the behavior which we observe comes from the recent work \cite{Amoretti:2018tzw}\footnote{We thank Blaise Gouteraux et al. for sharing their preliminary results with us.}. The authors take into account the extra hydrodynamical coefficient $\gamma_1$ which enters the Josephson relation multiplied by the second derivative of the chemical potential, and show that it modifies the expression for the AC conductivity in the pinned spontaneous case \eqref{AC model}\footnote{It is worth mentioning, however, that in \cite{Amoretti:2018tzw} the contribution from $\gamma_1$ turned out to be negligible, which may not be the case in our model.} 
 In our case, the measured small values of $\tilde{\Omega}$ (Figs.\,\ref{fig:fit_Mott_vsT}, \,\ref{fig:fit_Mott_vsLambda}) are such that $\tilde{\Omega}$ can safely be neglected in the denominator of \eqref{AC model} as compared to the $\Gamma$ and $\omega_0$ scales. 
 Therefore, in our setup the parameter $\gamma_1$ effectively adds an extra contribution to $\tilde{\Omega}$: 
\begin{equation}
\label{eq:shiftOmega}
\tilde{\Omega} = \Omega + \omega_0^2 \gamma_1.
\end{equation}
The coefficient $\gamma_1$ has been neglected in \cite{Delacretaz:2017zxd,Delacretaz:2016ivq} and has been also found to be negligible in \cite{Amoretti:2018tzw}. However in our case it appears to be playing an important role, since it naturally explains the observed $\lambda^2$ scaling of $\tilde{\Omega}$ (recall $\omega_0\sim \lambda$). 
Assuming that $\Omega$ does not depend on $\lambda$ it can be seen immediately from the fits of Fig.\,\ref{fig:OmegavsLambda} that $\gamma_1$ gives the leading contribution to $\tilde{\Omega}$. 
The presence a non-vanishing $\gamma_1$ deserves a deeper study, since it may indeed play a significant role 
in the conductivity of the pinned spontaneous insulator, and we plan to address it elsewhere.

\section{\label{sec:Conclusion}Conclusions}

In this work we study in detail the different gapless insulating states which one can encounter in holographic models. One class features explicit translation symmetry breaking, which is relevant in the IR and is the holographic counterpart of the conventional band insulator. 
The other class arises due to weak pining of spontaneous translational symmetry breaking and can be associated with 
the Mott insulators. We focus in particular on the physical mechanisms which govern the transport in these two classes.

We use the powerful technical tool developed recently in \cite{Donos:2018kkm}, which allows us to evaluate the incoherent conductivity at all frequencies and therefore to explore precisely the lineshapes of the AC conductivities in different classes of insulators.

We find that in the case of the explicit relevant insulator, the conductivity is completely determined by its incoherent part and the momentum mediated coherent transport is suppressed at low temperatures. Said differently, the incoherent conductivity cannot be understood in terms 
of the hydrodynamic low lying quasinormal modes. Instead it is set by the scaling features of the near horizon IR geometry.

On the other hand we show that the conductivity in the pinned spontaneous hologrpahic insulator is mediated by very different mechanism. 
In fact, we find that the incoherent conductivity only accounts for a small fraction of the total transport. Instead the leading contribution to the DC transport is coming from the shoulder of the gapped momentum transport peak. 
This is finite due to the nonzero coefficient $\tilde{\Omega}$, which was originally associated with the phase relaxation rate due to topological defects in \cite{Delacretaz:2017zxd}. However our finding demonstrates that the physical nature of this coefficient is different from the motion of topological defects in the spontaneous structure, suggested in \cite{Delacretaz:2017zxd}. 
More concretely, we find that it is proportional to the strength of the explicit pinning potential squared and therefore cannot be attributed to the features of pure spontaneous order parameter. The higher derivative hydrodynamic coefficient $\gamma_1$, considered recently in \cite{Amoretti:2018tzw}, can provide a more natural physical explanation to our results instead. However, 
within our current framework we cannot unambiguously isolate its value since we only measure its combination with the phase relaxation rate \eqref{eq:shiftOmega}.

Another explanation to the observed phenomenon may come from other mechanisms of phase relaxation, which may arise due to the explicit breaking of translations. The commensurate lock in and the creation of discommensurations \cite{Mott, Krikun:2017cyw} can destroy the space coherence of the spontaneous order parameter, resulting in the similar relaxation terms in the Josephson relation. The more precise study of $\gamma_1$ and its independent evaluation from the different observables will help to separate its contribution from the other potential mechanisms and will shed light on the nature of coherent transport in the pinned spontaneous holographic strange insulators.





\acknowledgments
We are especially grateful to Aristos Donos and Blaise Gouteraux who actively helped us setting up this project. 
We also thank Koenraad Schalm and Jan Zaanen for useful comments. We thank Blaise Gouteraux, Andrea Amoretti, 
Danielle Musso and Daniel Arean for sharing with us their preliminary results. We acknowledge the contributions 
of Nick Poovuttikul and Matteo Baggioli in the preliminary stages of this project.

A.K. wishes to thank the organizers and participants of the workshop {\em Many-body Quantum Chaos, 
Bad Metals and Holography} made possible by support from NORDITA, ICAM (the Institute for Complex 
Adaptive Matter) and Vetenskapradet, where the preliminary results of this project have been discussed.
A.K. also acknowledges the hospitality Barcelona University, where a considerable part of this project has been done.

The work of T.A. is supported by the ERC Advanced Grant GravBHs-692951.
The work of A.K. is supported by a Koenraad Schalm's VICI award of the Netherlands Organization for Scientific Research (NWO), by the Netherlands Organization for Scientific Research/Ministry of Science and Education (NWO/OCW), by the Foundation for Research into Fundamental Matter (FOM). 

\appendix

\section{\label{app:background}Numerical background solutions}

We construct all solutions numerically following \cite{Andrade:2017cnc}. In particular we use the DeTurk
trick to fix the gauge symmetry of the metric ansatz, and solve the resulting non-linear
equations by a Newton-Raphson procedure. 
In all our calculations, we have used a homogeneous grid of 80 points in the radial direction. 
The near boundary asymptotics of the background fields read
\begin{align}
\label{UV bckg 1}
	U &= 1 + u^2 U^{(2)} + u^4 U^{(4)}  + u^4  \log u \tilde U^{(4)}   + \ldots \\
	T &= 1 + u^2 T^{(2)} + u^4 T^{(4)}  + u^4  \log u \tilde T^{(4)}   +  \ldots  \\
	W_i &= W_i^{(0)} + u^2 W_i^{(2)} + u^4 W_i^{(4)}  + u^4  \log u \tilde W_i^{(4)}   +  \ldots   \\
	Q &=  u^4 Q^{(4)} + u^4  \log u \tilde Q^{(4)}  \ldots \\
	A_t &= \mu + u^2  A_t^{(2)} + u^2  \log u \tilde A_t^{(2)}  + \ldots \\
 	A_2 &=  u^2  A_2^{(2)} + u^2  \log u \tilde A_2^{(2)}   + \ldots \\	
	B_t &= u^2  B_t^{(2)} + u^2  \log u \tilde B_t^{(2)}  \\
\label{UV bckg 2}
 	B_2 &=  \lambda + u^2  B_2^{(2)} +  u^2  \log u \tilde B_2^{(2)}   + \ldots 
\end{align}

The near horizon asymptotics are given by
\begin{align}
\label{IR bckg 1}
	U &= U_H^{(0)} + (1-u) U_H^{(1)} + \ldots \\
	T &= T_H^{(0)} + (1-u) T_H^{(1)} + \ldots \\
	W_i &= W_{i,H}^{(0)} + (1-u) W_{i,H}^{(1)} + \ldots \\
	Q &= (1-u) [ Q_H^{(0)} + (1-u) Q_H^{(1)} ] + \ldots \\
	A_t &= (1-u) [ A_{t,H}^{(0)} + (1-u) A_{t,H}^{(1)} ] + \ldots \\
 	A_2 &=  A_{2,H}^{(0)} + (1-u) A_{2,H}^{(1)}  + \ldots \\	
	B_t &= (1-u) [ B_{t,H}^{(0)} + (1-u) B_{t,H}^{(1)} ] + \ldots \\
\label{IR bckg 2}
 	B_2 &=  B_{2,H}^{(0)} + (1-u) B_{2,H}^{(1)}  + \ldots 
\end{align}

\section{\label{app:zero_T}Zero temperature analysis}

The behavior of the background profiles near horizon at zero temperature can by analyzed by studying the near horizon expansion of the equations of motion following from \eqref{eq:action}, what we do in complete analogy to the treatment of \cite{Donos:2012js}. At zero temperature the horizon of the black hole gets critical and the $g_tt$ component of the metric develops a double zero. Therefore instead of \eqref{IR bckg 1} we use an ansatz with $U_H^{(0)} \rar 0$. We assume the arbitrary leading exponents in the behavior of the other fields
\begin{align}
 W_1 &\sim W_1^0 r^{\alpha_{W_1}}, & W_2 &\sim W_2^0 r^{\alpha_{W_2}}, & W_3 &\sim W_3^0 r^{\alpha_{W_3}} \\
 \notag
 A_t &\sim  A_t^0 r^{\alpha_{A_t}}, & A_2 &= b^0 + b^1 r^{\alpha_{A_2}}, & & \\
\notag
 B_2 &\sim w^0 + w^1 r^{\alpha_{B_2}}, &   Q &= Q^0 r^{\alpha_{Q}}, 
\end{align}
and obtain the powers $\alpha$ from the consistency conditions of the leading order near horizon expansions for the equations of motion. This procedure gives several solutions corresponding to different zero temperature ground states, including \eqref{eq:zeroT_explicit} and \eqref{eq:zeroT_spont}.

The irrelevant deformations of this near horizon expansion can be obtained by the analysis of the linearized perturbations around the ground state solutions. Similarly to the background we assume the arbitrary exponents of the leading power law expansions for the linearized perturbations and these powers are obtained by solving the consistency condition of the linear system resulting from the near horizon expansion of the equations for perturbations. We find the spectrum of deformation exactly analogous to \cite{Donos:2012js}. In the spontaneous case we find an additional mode due to weak pinning \eqref{eq:lambda_deforamtion}. We checked that the irrelevant deformation modes never acquire the imaginary powers and therefore the ground states which we discuss are stable IR fixed points.

One of the marginal deformations is particularly important since it accounts to moving away from the zero temperature state. This is the deformation of $U$-field of order $1$ at the horizon. One can check that it is equivalent to the shift in coordinate $r$ applied to all the field profiles. Therefore the expressions for infinitesimal temperature \eqref{eq:explicit_sig_scaling} and \eqref{eq:spont_sig_scaling} can be obtained by evaluating the zero temperature profiles at the distance $r\sim T$ from the critical horizon.


\section{\label{app:DC_formulas}DC conductivities from near horizon data}

Following \cite{Donos:2014cya, Donos:2015bxe,Banks:2015wha,Donos:2015gia,Donos:2017mhp}, 
we can find expressions for the DC limit of the thermoelectric conductivities. 
The first step is to identify the bulk radially conserved currents that yield the boundary 
currents when pulled back to the boundary. 
As we have already mentioned, there are no extra contributions due to the 
magnetization currents, which largely simplifies our analysis. 

The electric current easily follows from the equation of motion for the gauge field $A$, 
which we can write as
\begin{equation}
	\partial_\mu H^{\mu \nu} = 0
\end{equation}
\noindent where
%
%
\begin{equation}
	H^{\mu \nu} =  F^{\mu \nu} -  \frac{\gamma}{4} \epsilon^{\mu \nu \alpha \beta \gamma} A_\alpha F_{\beta \gamma} + 
	 \kappa \epsilon^{\mu \nu \alpha \beta \gamma} B_\alpha W_{\beta \gamma}  
\end{equation}
\noindent This allows us to define the boundary current
\begin{equation}
\label{Ji}
	J^i = \sqrt{-g} H^{u i}
\end{equation}
It is straightforward to check that the linearized current $\delta J^x$ defined
as the variation of \eqref{Ji} satisfies $\partial_u \delta J^x = 0$ as a consequence
of the linearized equations of motion. 
As we shall see below, this corresponds precisely to the boundary current 
when evaluated at $u=0$, so it is indeed the quantity we are after. 

In order to construct the heat current, we consider
\begin{equation}
	G^{\mu \nu} = 2 \nabla^{[\mu} k^{\nu]} - \frac{2}{3} k^{[\mu} F^{\nu] \rho} A_\rho 
	- \frac{2}{3} k^{[\mu} W^{\nu] \rho} B_\rho - A_t H^{\mu \nu} - B_t \tilde H^{\mu \nu} 
\end{equation}
\noindent where
\begin{equation}
	\tilde H^{\mu \nu} = W^{\mu \nu} + 2 \kappa \epsilon^{\mu \nu \alpha \beta \gamma} A_\alpha W_{\beta \gamma}
\end{equation}
\noindent and $k = \partial_t$. Following the calculations in 
\cite{Donos:2014cya, Donos:2015bxe,Banks:2015wha,Donos:2015gia,Donos:2017mhp}, we can show that in general, 
this quantity leads to a conserved current,
\begin{equation}
\label{Qi}
	Q^i = \sqrt{-g} G^{u i}
\end{equation}
Indeed, it is straightforward to verify that $\partial_u \delta Q^x = 0$ by virtue of the 
linearized equations of motion. The pull-back of $Q^x$ at the boundary leads to the desired
boundary operator, as we shall see below. 

\begin{figure}[th]
\centering
\includegraphics[width=1 \linewidth]{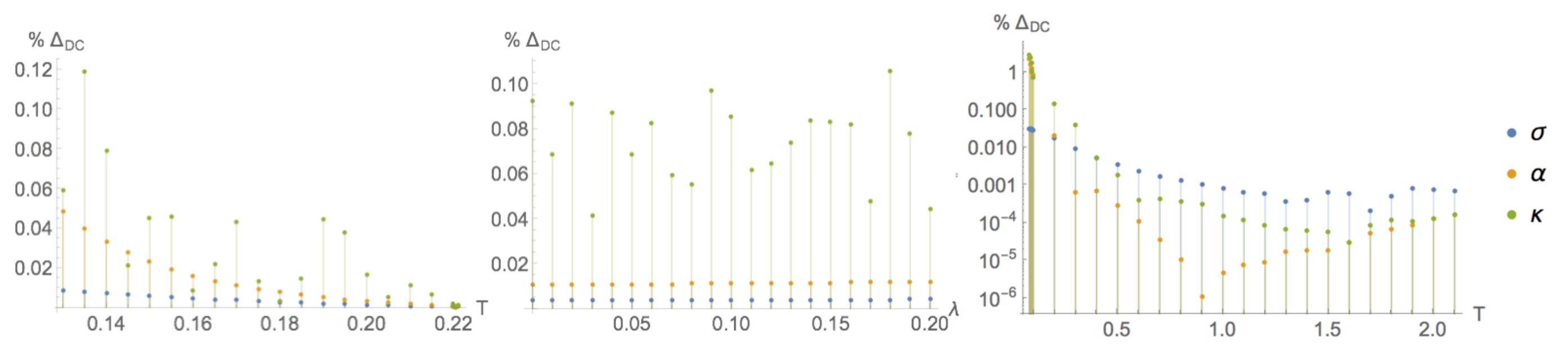}
\caption{Percentage difference between DC calculation from horizon data and AC calculation at
$\omega = 10^{-8}$. (Left): Pinned spontaneous data \eqref{eq:pinned spontanoues parameters} at fixed $\lambda = 0.1$; (Center): Pinned spontaneous data \eqref{eq:pinned spontanoues parameters} at fixed $T/T_c = 0.76$; (Right): explicit insulator \eqref{eq:explicit parameters} at fixed $\lambda = 3$.}%
\label{fig:DeltaSigmaDC}
\end{figure}

The next step is to consider a set of perturbations with a particular linear in time
dependence which allows us to introduce DC sources in a convenient way. 
This can be found in \cite{Donos:2015bxe}, and for the present case reduce to 
\begin{align}
\nonumber
	\delta (ds^2) &= 2[ (\delta g_{u1} du + \delta g_{t1} dt) \omega_1 + (\delta g_{u3} du + \delta g_{t3} dt) \omega_3 +
	(\delta g_{12} \omega_1 + \delta g_{23} \omega_3) \omega_2]  \\
	& + 2 t \zeta \omega_1 ( g_{tt} dt + g_{t2} \omega_2 ) \\
	\delta A &= \delta A_1 \omega_1 + \delta A_3 \omega_3 + t ( - E + A_t \zeta) \\
	\delta B &= \delta B_1 \omega_1 + \delta B_3 \omega_3 + t  B_t \zeta
\end{align}
\noindent where $E$ and $\zeta$ are constant, while all the other unknowns are functions 
of the radial coordinate. 
As argued in \cite{Donos:2015bxe}, this corresponds to a particular diffeomorphism plus a $U(1)$ gauge
transformation, guaranteeing that the time dependence drops out of the equations 
of motion, as we have explicitly checked.
When evaluated at the boundary, the perturbed radial currents $\delta J^x$, $\delta Q^x$ reduce to 
the expressions for the dual operators \eqref{Jx UV}, \eqref{Qx UV}.
Therefore, using the fact that $\partial_u \delta J^x=\partial_u \delta Q^x=0$, 
we can obtain expressions for the currents in terms of the sources $E_x$, $\zeta_x$,
and the horizon data. These can be written in terms of the coefficients
of the near horizon expansions \eqref{IR bckg 1}-\eqref{IR bckg 2}, as 
\begin{align}
\label{JH}
	\delta J^x &= (-det g_H)^{-1/2} W_{2,H}^{(0)} [ U_H^{(0)}  W_{3,H}^{(0)} E - A_{t,H}^{(0)} W_{3,H}^{(0)} \delta g_{t1}^{(0)} 
	- k U_H^{(0)}  A_{t,H}^{(0)}  \delta g_{t3}^{(0)}   ] \\
\label{QH}
	\delta Q^x &= - (4 \pi T) (-det g_H)^{-1/2}    U_H^{(0)}  W_{2,H}^{(0)}  W_{3,H}^{(0)} \delta g_{t1}^{(0)} 
\end{align}
\noindent where $(-det g_H) =  U_H^{(0)}  T_H^{(0)} W_{1,H}^{(0)} W_{2,H}^{(0)}  W_{3,H}^{(0)}  $
The linearized pieces of horizon data $\delta g_{t1}^{(0)}$, $\delta g_{t3}^{(0)}$ can be eliminated from 
\eqref{JH}, \eqref{QH} using the constraint equations evaluated at the horizon. 
%
Evaluating the constraint equations at the horizon, we obtain the relations,
\begin{align}
\label{to solve1}
	-(4 \pi T) \zeta - \frac{ A_{t,H}^{(0)} }{ T_H^{(0)} } E + C_1 \delta g_{t1}^{(0)}  + C_2 \delta g_{t3}^{(0)} &= 0 \\
	\frac{ B_{t,H}^{(0)}  } { W_{1,H}^{(0)} } E + C_3 \delta g_{t1}^{(0)}  + C_4 \delta g_{t3}^{(0)} &= 0 
\end{align}
\noindent where
\begin{align}
	C_1 &= \frac{k^2}{ W_{1,H}^{(0)} W_{2,H}^{(0)} W_{3,H}^{(0)}  }  [ W_{2,H}^{(0)}  (A_{3,H}^{(0)} )^2 + 
	(W_{2,H}^{(0)} )^2+ (W_{3,H}^{(0)} )^2 + + W_{2,H}^{(0)} ( (B_{3,H}^{(0)} )^2 - 2 W_{3,H}^{(0)}  ] \\
	C_2 &=\frac{k}{ T_{H}^{(0)}  W_{3,H}^{(0)} } [ A_{t,H}^{(0)} A_{3,H}^{(0)} + Q_{H}^{(0)} W_{2,H}^{(0)}  + B_{t,H}^{(0)} B_{3,H}^{(0)}  ] \\
	C_3 &= \frac{1}{T_{H}^{(0)} W_{1,H}^{(0)} } [ A_{t,H}^{(0)} A_{3,H}^{(0)} + Q_{H}^{(0)} W_{2,H}^{(0)}  + B_{t,H}^{(0)} B_{3,H}^{(0)}  ] \\
\label{to solve6}
	C_4 &= \frac{k}{W_{1,H}^{(0)}  W_{3,H}^{(0)}} [ (A_{3,H}^{(0)} )^2 +  (B_{3,H}^{(0)} )^2 + W_{2,H}^{(0)}    ]
\end{align}
It is clear that using equations \eqref{to solve1}-\eqref{to solve6} we can write the currents in terms of the sources 
and the horizon data of the background fields, as desired. The final expressions are not illuminating, so we 
do not transcribe them.
The elements of the DC thermoelectric
conductivities are then simply given by the partial derivatives with respect to the sources as
\begin{equation}
\label{equ:DC_formulae}
	\sigma_{DC} = \partial_E \delta j^x, \qquad \alpha_{DC} = \partial_\zeta \delta j^x, \qquad 
	\bar \alpha_{DC} = \partial_E \delta Q^x, \qquad \bar \kappa_{DC} = \partial_\zeta Q^x
\end{equation}
We can readily check that the Onsager relation $\alpha_{DC} = \bar \alpha_{DC} $ hods, 
providing a non-trivial check of our procedure. Moreover, we find excellent agreement with the 
small frequency AC calculation, see Fig. \ref{fig:DeltaSigmaDC}. The main source of inaccuracy 
is the computation of the thermal conductivity $\bar \kappa$, since it involves extracting a term which 
is suppressed by $u^4$ in the linearized numerics for the AC calculation. 

\section{\label{app:AC_conductivities}AC conductivities from linearized equations of motion}

In this appendix we provide a more detailed description of our computation of the frequency dependence
thermoelectric conductivity. 
As mentioned in the main text, the sources of interest are contained in the UV expansion of the fields 
$\delta g_{t1}$, $\delta A_1$, see \eqref{pert g A}. All fields are assume to depend on the 
radial coordinate $u$ and to have a harmonic time dependence $e^{- i \omega t}$. 
In order to find a consistent system of perturbation equations, we need to consider the fluctuations
\begin{equation}
 \{	\delta g_{u1}, \delta g_{u3}, \delta g_{t1}, \delta g_{t3}, \delta g_{12} , \delta g_{23}, 
 \delta A_1, \delta A_3, \delta B_1, \delta B_3 \}
\end{equation}
This set of perturbations contains gauge ambiguities which we fix using the DeDonder gauge 
as in \cite{Rangamani:2015hka}. More specifically, denoting by $\delta E_{\mu \nu}$ to the 
linearized Einstein equations, we replace them by $\delta E_{\mu \nu} \to 
\delta E_{\mu \nu} + \tau_{\mu \nu}$ where the gauge fixing term 
corresponds to 
\begin{equation}
	\tau_{\mu \nu} = \nabla_{(\mu} \tau_{\nu)}, \qquad \tau_\mu = \nabla^\nu \delta g_{\mu \nu}
\end{equation}
Solving the asymptotic equations near $u=0$, we obtain
\begin{align}
\label{UV lin}
	\delta g_{a b} &= u^{-2} ( \delta g_{ab}^{(0)} + u^2 \delta g_{ab}^{(2)}  + u^4 \delta g_{ab}^{(4)} 
	+ u^4 \log u \delta \tilde g_{ab}^{(4)}  + \ldots   ) \\
	 \delta g_{u a} &=u^{-1} ( \delta g_{ua}^{(0)} + u^2 \delta g_{ua}^{(2)}  + u^4 \delta g_{ua}^{(4)} 
	+ u^4 \log u \delta \tilde g_{ua}^{(4)}  + \ldots   ) 
\end{align}
\noindent where $x^a = {t, x^i}$. 
Near the horizon we find that regular perturbations must satisfy 
\begin{align}
\label{IR lin}
	\delta g_{a b} &= (1-u)^\alpha ( \delta g^H_{ab} + O[ (1-u)]  ) \\
	 \delta g_{u a} &=(1-u)^{\alpha-1} ( \delta g^H_{ab} + O[ (1-u)]  )
\end{align}
\noindent where $\alpha = - i \omega/(4 \pi T)$. 
We solve the resulting perturbation equations imposing ingoing 
boundary conditions at the horizon, and fixing the sources on the UV by means 
of the asymptotic expansions \eqref{UV lin}, \eqref{IR lin}. 
We also obtain the QNMs by setting all sources to zero. 

According to the standard AdS/CFT dictionary, the dual operators are encoded in the subleading 
terms $\delta g_{tx}^{(4)}$, $\delta A^{(2)}_x$. Using the counter-term prescription of \cite{Erdmenger:2015qqa} 
we arrive at
\begin{align}
\label{Jx UV}
	 J^x  &= 2 \delta A^{(2)}_x  - \rho \delta g_{tx}^{(0)} \\
\label{Qx UV}
	 Q^x  &= 4 \delta g_{tx}^{(4)} + 2 \mu \delta A^{(2)}_x +\left( 2 A^{(2)}_t + 4 T^{(4)} + 
	\frac{2}{3} \mu^2 + 4\right) \delta g_{tx}^{(0)}
\end{align}

\bibliographystyle{JHEP-2}
\bibliography{strange_insulators}

\end{document}